\documentstyle[psfig,epsf]{l-aa}
\def\HI{H\,{\sc i}\ }
\def\HII{H\,{\sc ii}\ }
\def\pheins{\phantom{1}}

\def\phdoteins{\phantom{.1}}
\def\ROSAT{{\it ROSAT\ }}
\def\EINSTEIN{{\it Einstein\ }}
\def\GINGA{{\it Ginga\ }}
\begin{document}

   \thesaurus{11.09.1 M83; 11.08.1; 11.09.4; 09.07.1; 09.13.1; 13.25.2}
  \title{ROSAT PSPC X-ray observations of the nearby spiral galaxy M83}
  \subtitle{}
  \author{M. Ehle\inst{1}$^,$\inst{2}$^,$\inst{3}, W. Pietsch\inst{1}, 
  R. Beck\inst{2}, U. Klein\inst{4}}
  \offprints{M. Ehle, mehle@mpe.mpg.de}
  \institute{Max-Planck-Institut f\"ur Extraterrestrische Physik, 
	     Giessenbachstra{\ss}e, 85740 Garching, Germany
             \and
             Max-Planck-Institut f\"ur Radioastronomie,
             Auf dem H\"ugel 69, 53121 Bonn, Germany
	     \and
             Australia Telescope National Facility, P.O. Box 76, 
	     Epping N.S.W. 2121, Australia	     
	     \and
	     Radioastronomisches Institut der Universit\"at Bonn, Auf
	     dem H\"ugel 71, 53121 Bonn, Germany             
             }
  \date{Received 23 October 1995/ Accepted 9 June 1997}
  \maketitle
  \markboth{M. Ehle et al.: ROSAT PSPC X-ray observations of the nearby spiral galaxy M83}
	   {M. Ehle et al.: ROSAT PSPC X-ray observations of the nearby spiral galaxy M83}
\begin{abstract}
{
The nearly face-on SBc galaxy M83 (NGC~5236) was observed for 25 ksec 
with the \ROSAT PSPC. We detected 13 point-like sources in this galaxy,
10 of which were previously unknown, down to a limiting
luminosity of $\sim0.8\cdot10^{38}$ erg s$^{-1}$ (for D=8.9~Mpc).
Eight of these sources are positionally associated with \HII regions
and/or \HI voids, suggesting an association with the younger stellar
population, variable X-ray binaries or with hot expanding gaseous bubbles.

We measured extended X-ray radiation from almost the whole optically visible
galaxy with a luminosity of $L_{\rm x}\sim3.6\cdot10^{40}$~erg s$^{-1}$ 
in the energy range 0.1-2.4~keV. Approximately $20\%$ of this 
emission can be explained by undetected point-like sources. 
We detected diffuse soft (0.1-0.4~keV) X-ray emission due to 
hot gas with a luminosity of 
$L_{\rm x}{\rm(soft)}\sim1.4\cdot10^{40}$~erg s$^{-1}$.
Comparing the diffuse soft and hard X-ray emission components, we 
observed a different asymmetric distribution and a slower radial decrease 
of the intensity profile of the soft X-ray emission. Both these results 
support the existence of a huge spherical gas halo of $\sim10-15$~kpc radius. 
On the other hand, the radial scale lengths of the hard X-ray radiation, 
that of the thermal radio emission and the profile of the optical surface 
brightness are similar, favouring the idea that all these emission 
processes are connected to star formation in the galaxy's disk. 

M83 is the first face-on galaxy where the diffuse X-ray emission spectrum 
can be characterized by a two-temperature thermal plasma: a soft X-ray 
emitting warm `halo component' of plasma temperature $2.1\cdot10^6$~K and an 
internally absorbed hot `disk component' at $5.7\cdot10^6$~K which is
dominating the emission in the hard (0.5-2.0 keV) \ROSAT energy range.
The (distance-independent) mean surface brightness of the soft diffuse 
emission of M83 is about twice that of NGC~253.

The electron densities in the galactic halo of M83 ($\sim
1.2\cdot10^{-3}/\sqrt{\eta}$~cm$^{-3}$, with a volume filling
factor $\eta$) are in general too low to explain the observed
depolarization of the radio emission. This indicates that the
ionization equilibrium might be violated, leading to higher electron
densities.
The combination of X-ray and radio polarization observations allows
an estimate of the plasma parameter $\beta = U_{\rm therm}/U_{\rm magn}$
which is found to be $0.2\pm0.1$.
This result supports the hypothesis that magnetic fields play an 
important role for the evolution and structure of galactic gas haloes.
The high energy input rate in the active star-forming disk of M83 seems 
to be responsible for the outflow of hot gas and the halo formation.

}
  \end{abstract}
%
%
\section{Introduction}
\label{intro}
The nearby spiral galaxy M83 (NGC~5236) is a classical grand-design spiral
located at a distance of 8.9 Mpc\footnote{Distance determinations of M83
have come up with rather different values; here we adopt that of 
Sandage \& Tammann (1975) and scale all luminosities to that value}.
It has been the target of numerous astrophysical investigations in various 
spectral regimes because of its proximity. 
Table~\ref{parameters} lists some general parameters of M83.

\begin{table*}
\caption{Parameters for M83 (NGC~5236)}
\label{parameters}
\begin{flushleft}
\begin{tabular}{lll}
\hline
\noalign{\smallskip}
Parameter & Value & Reference and Comments\\
\noalign{\smallskip}
\hline
\noalign{\smallskip}
Type               & SBc(s)II   &  Sandage \& Tammann (1987; RSA)   \\
R.A. (J2000)      & $13^{\rm h}37^{\rm m}0\fs25$  & Condon et al. (1982), central radio source\\
Decl. (J2000)     & $-29\degr51\arcmin51\farcs3$  & Condon et al. (1982), central radio source\\
Position angle     &            &      \\
of major axis      & $45\degr$  &  Talbot et al. (1979)   \\
Inclination        & $24\degr$  &  Talbot et al. (1979)   \\
Distance           & 8.9 Mpc       &  Sandage and Tammann (1975)\\
		   &               &  $10\arcsec\cor0.43$ kpc\\
Size ($D_{25}$)    & 29 kpc        &  de Vaucouleurs et al. (1976)   \\
Total mass         & $1 - 5\cdot10^{11} M_{\odot}$& Huchtmeier \& Bohnenstengel (1981),\\ 
		   &                              & derived from rotation curve\\
\HI mass            & $2.4\cdot10^{10} M_{\odot}$  & Huchtmeier \& Bohnenstengel (1981),\\
		   &                              & integrated over the extent of the \HI distribution ($76\arcmin\times94\farcm6$)\\
\noalign{\smallskip}
\hline
\noalign{\smallskip}
Abs. blue magnitude& -21.7 & RSA Catalog of Bright Galaxies\\
		   & $\cor2.8\cdot10^{44}$ erg/s &         \\
X-ray luminosity   & $3.2\cdot10^{40}$ erg/s & \EINSTEIN IPC (0.5-3.0)~keV, Trinchieri et al. (1985)\\
                   & $4.2\cdot10^{40}$ erg/s & \EINSTEIN HRI (0.5-3.0)~keV, Trinchieri et al. (1985)\\
		   & $4.5\cdot10^{40}$ erg/s & \GINGA (2-10)~keV, Ohashi et al. (1990)\\
		   & $5.7\cdot10^{40}$ erg/s & \ROSAT PSPC (0.1-2.4)~keV, this work\\
\noalign{\smallskip}
\hline
\end{tabular}
\end{flushleft}
\end{table*}

M83 was first observed in X-rays with the \EINSTEIN observatory Imaging
Proportional Counter (IPC) as part of a sample of nearby spiral galaxies
to study their integrated properties (Fabbiano et al. 1984). 
Trinchieri et al. (1985) obtained high-resolution X-ray observations with 
the \EINSTEIN HRI and detected three unresolved sources 
and a nuclear component. 

The starburst nucleus of M83 is a strong X-ray source that was detected
in the \EINSTEIN observations with a luminosity of
$L_{\rm x}\sim1\cdot10^{40}$ erg s$^{-1}$.
Trinchieri et al. (1985) suggested that the X-ray emission
from the nucleus is not dominated by a contribution of massive binary
star systems, but is due to a less evolved X-ray emitting population.
They also discussed the possibility that a large 
fraction of the X-ray emission from the nucleus of M83 could be produced 
by thermal emission from hot shock-heated gas ejected from the nucleus, 
as is detected in NGC\,1068 (Wilson et al. 1992), M82 (Watson 
et al. 1984) and NGC\,253 (Fabbiano \& Trinchieri 1984, Fabbiano 1988, 
Pietsch 1993). A further peculiarity of the nuclear region of M83 is the
existence of a polar, circum-nucleus dust lane probably in a plane tilted 
from the galactic plane by about $70\degr$ (Sofue \& Wakamatsu 1994). 
Reducing the accretion to the nucleus, this ring structure might affect 
the nuclear star-forming activity.

The \EINSTEIN IPC observations indicated an X-ray extent 
comparable in size and shape to that seen in blue light. 
This emission component was discussed as probably due to faint 
individual sources (evolved stellar population, LMXBs), 
distributed uniformly across the disk of the galaxy
(Trinchieri et al. 1985). 

Ohashi et al. (1990) observed M83 in the harder X-ray energy band (2-20 keV) 
using the \GINGA satellite. The observed spectra are similar to those of 
low-mass X-ray binaries (LMXBs) but significantly softer than those of 
X-ray pulsars.
The spectra can also be described by a thermal bremsstrahlung model with a 
temperature of $\sim8\cdot10^7$ K, suggesting that a major fraction of the 
X-ray emission may originate from very hot gas. As the time scale for
starburst activity is significantly shorter than the typical lifetime of
LMXBs, Ohashi et al. find it unlikely that the high X-ray luminosity
of M83 is caused by LMXBs and instead favour the idea that hot gas (both 
within and outside of SNRs) produces their observed X-ray spectra.

In this paper we report X-ray observations of M83 carried out with the
\ROSAT X-ray telescope (Tr\"umper 1983, Aschenbach 1988) Position 
Sensitive Proportional Counter (PSPC, Pfeffermann et al. 1987).
The foreground Galactic X-ray-absorbing gas is distributed smoothly across the
optical extent of M83 (Hartmann \& Burton 1996) and has a relatively low
column density ($3.97\cdot10^{20}$ cm$^{-2}$; Dickey \& Lockman
1990). Hence our observations allow the detection of diffuse soft
(0.1-0.4~keV) X-ray emission due to hot gas in M83.
In Section 2 we give the observational details and data analysis. In Section 
3 we discuss the different emission components and investigate the 
interaction of hot gas, galactic winds and magnetic fields. 
Finally, we compare M83 with other nearby galaxies where diffuse X-ray
emission was detected.

\section{Observations and data analysis}
\label{obs}
The data analysis and reduction was performed using the EXtended Scientific 
Analysis System EXSAS (Zimmermann et al. 1994). 

The field of view of the \ROSAT PSPC detector was centered at the 
pointing position RA $=13^{\rm h}37^{\rm m}00\fs0$, 
Dec $=-29\degr52\arcmin12\farcs0$ (J2000) so that the X-ray emission of 
M83 was contained well within the innermost 6\arcmin~ from the center.
A total of 16 PSPC pointings in the period from Jan 1992 to Jan 1993 (see
Table~\ref{schedule}) resulted in a 25 ksec integration time.

\begin{table}
\caption{\ROSAT PSPC observations of M83}
\label{schedule}
\begin{flushleft}
\begin{tabular}{ccr}
\hline
\noalign{\smallskip}
Observing      & Date of     & Time\\
interval (OBI) & observation & (s)     \\
\noalign{\smallskip}
\hline
\noalign{\smallskip}
01 & 1992 Jan 28 & 644\\
02 & 1992 Aug 09 & 1363\\
04-08 & 1993 Jan 08-09 & 8291\\
09,11-18 & 1993 Jan 10-11 & 15073\\
\noalign{\smallskip}
\hline
\end{tabular}
\end{flushleft}
\end{table}

To investigate the data quality and positional offsets, we produced maps 
of the X-ray emission for each observing interval separately. Comparing the 
X-ray backgrounds in the individual maps, we looked for contamination due to
solar and auroral X-ray photons or due to particles of the cosmic
background emission (Snowden \& Freyberg 1993, Snowden et al. 1994).
Selection of data with low background values ($<$ 20 cts per 60
sec interval) reduced the finally analysed time to 21 ksec 
but decreased the background noise by $\sim20\%$. 

We looked for positional offsets between X-ray sources in the 
individual maps and found them to be less than $\sim\pm10\arcsec$. 
Systematical shifts between the maps could not be found. 

The absolute position of the field depends on the accuracy of
the PSPC pointing and has been checked comparing point-like X-ray
emission features with the positions of stars or extragalactic sources. 
Checking the positions of stars in the Hubble Space Telescope Guide Star 
Catalog no identification was obtained. Using the extragalactic database
NED \footnote{The NASA/IPAC extragalactic database (NED) is operated by the 
Jet Propulsion Laboratory, Caltech, under contract with the National 
Aeronautics and Space Administration} yielded two identifications: 
the X-ray source at RA $\sim13^{\rm h}38^{\rm m}43^{\rm s}$,
Dec $\sim-29\degr33\farcm5$ (J2000) is at the position of the F5V star 
HR\,5128 (SAO\,181825) which was already detected in the \EINSTEIN
survey (Schmitt et al. 1985). A second identification may exist for 
X-ray Source 14 (see Table~\ref{master}) and the quasar MS1332.6-2935.
However, the positional uncertainties of these {\it ROSAT}-detected 
sources are large because of their large off-axis positions (29\farcm0 
and 26\farcm6, respectively). 
We could identify several \ROSAT X-ray sources in the
field of our observations with faint blue stellar objects on ``ESO
Quick Blue" plates. Details of source identifications will be given in
Section \ref{sources}. No systematic positional offsets could be
found, and we decided not to apply any ``boresight" corrections for the
analysis reported in this paper.

\subsection{Derivation of X-ray images}
\label{maps}
Binning the data to a pixelsize of 5\arcsec~ we produced images 
of the inner 40\arcmin~ region of the PSPC detector
window. These have been exposure-corrected and smoothed in 8 energy 
subbands (0.11-0.19, 0.20-0.41, 0.42-0.51, 0.52-0.69, 0.70-0.90, 
0.91-1.31, 1.32-2.01 and 2.02-2.35~keV). 
The smoothing was done with a Gaussian filter with full width at half
maximum (FWHM) corresponding to the average resolution of the point 
spread function (PSF) at the detector center in the individual energy 
subbands. Then the subbands were merged again to form five 
standard energy bands (cp. Vogler et al. 1996). 
Table~\ref{baender} gives the selected \ROSAT energy
bands, FWHMs of the Gaussians used for smoothing, sigma (standard deviation)
and background values of the maps computed in `source-free' regions. (We note
that - as our observations are photon limited - sigma does not give the 
real background noise in the maps).
\begin{table*}
\caption[]{Characteristics of the M83 maps}
\label{baender}
\begin{flushleft}
\begin{tabular}{lllccc}
\hline
\noalign{\smallskip}
\multicolumn{3}{l}{\ROSAT energy band} & FWHM of Gaussians & sigma & Background\\ &      &        & for smoothing (\arcsec) & \multicolumn{2}{c}{$(10^{-6}$ cts s$^{-1}$ arcmin$^{-2})$}\\
\noalign{\smallskip}
\hline
\noalign{\smallskip}
Broad& B & (0.1-2.4 keV) & 52, 38, 28, 25 & 308 & 1440\\
Soft & S & (0.1-0.4 keV) & 52, 38 & 190 & 1002\\
Hard & H & (0.5-2.0 keV) & 28, 25 & 221 & 386\\
Hard1& H1& (0.5-0.9 keV) & 28, 25 & 149 & 258\\
Hard2& H2& (0.9-2.0 keV) & 25     & 151 & 138\\
\noalign{\smallskip}
\hline
\end{tabular}
\end{flushleft}
\end{table*}

Figure~1 shows a contour plot of the broad-band emission of
the whole inner 40\arcmin~ field-of-view of the PSPC. Around the extended
emission of M83, several previously unknown X-ray point-sources 
and diffuse emission from a cluster of galaxies in the
north-east are visible.
\begin{figure*}
{\centering\psfig{figure=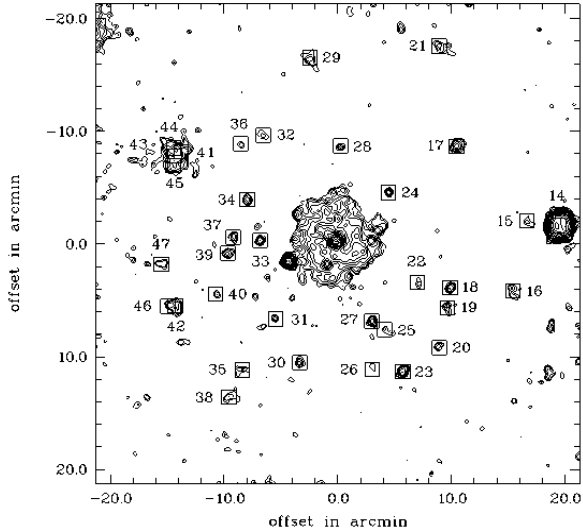,width=120mm,angle=-90}}
\caption[]{Contour plot of the broad-band \ROSAT PSPC image of the inner
40\arcmin~ of the M83 field. The image has been combined from images
in the individual energy bands which have been smoothed with a Gaussian
filter, with the FWHM of the average PSPC point response function and 
exposure-corrected for the respective energy bands. Contour levels
increase by $2^{n/2}$. Contours are $(1,1.4,2,2.8,4,...)\times1\cdot10^{-3}$
cts s$^{-1}$ arcmin$^{-2}$ above the background (FWHM and 
background values are given in Table~\ref{baender}). {\it ROSAT}-detected
sources outside of M83 are plotted as squares, with \ROSAT source numbers. 
Offsets are with respect to the center of the
observation RA $=13^{\rm h}37^{\rm m}00\fs0$,
Dec $=-29\degr52\arcmin12\farcs0$ (J2000). North is at top, and west to 
the right}
\label{field}
\end{figure*}

\begin{figure*}
{\centering\psfig{figure=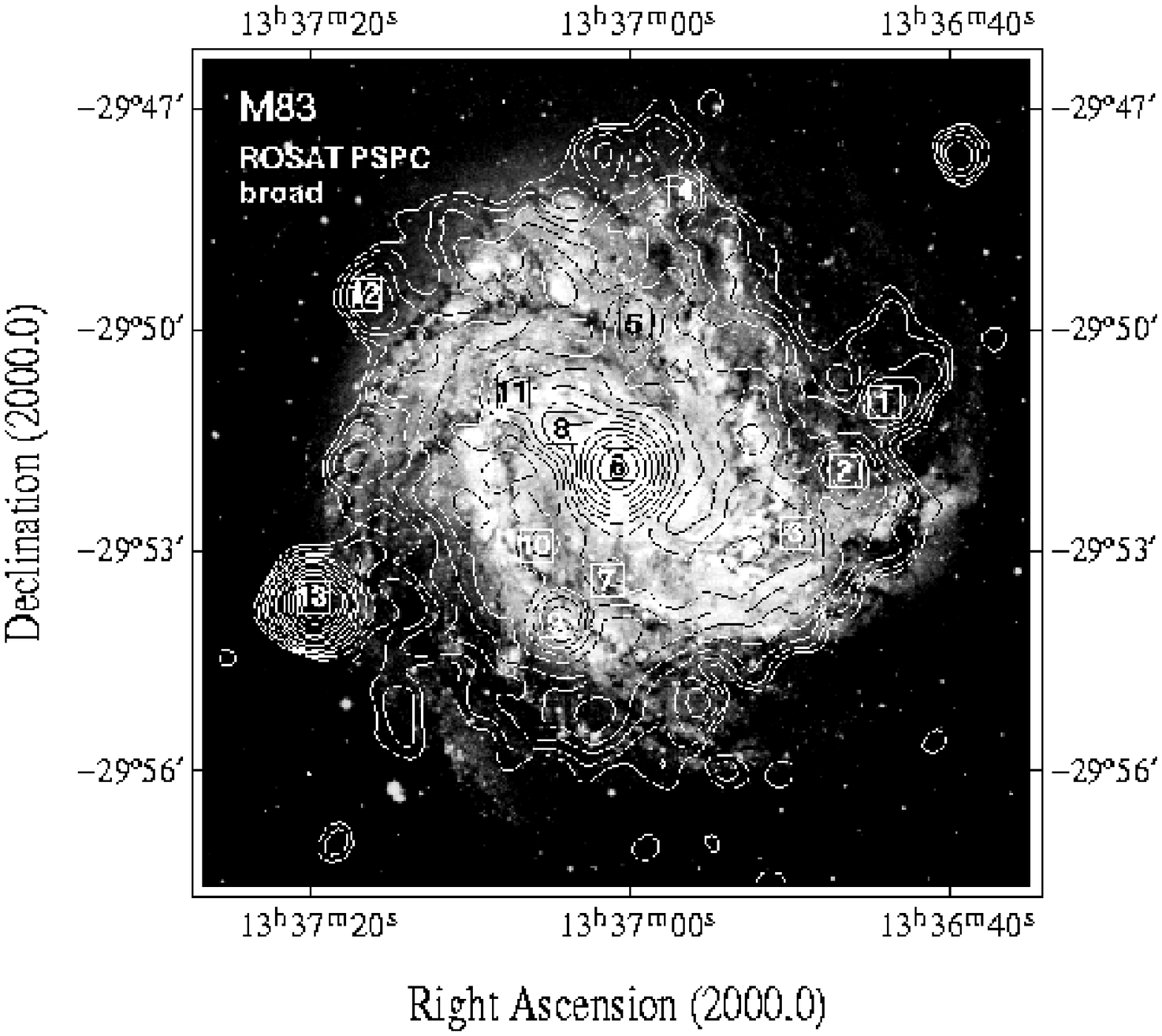,width=150mm,bbllx=25pt,bblly=180pt,bburx=569pt,bbury=661pt}}
\caption[]{Broad-band \ROSAT PSPC contours as in Fig.~\ref{field} overlaid 
onto a deep optical image from the 3.9m Anglo Australian Telescope, kindly 
provided by David Malin. 
Here we are showing only the central part of the PSPC field of view.
{\it ROSAT}-detected sources in M83 are marked as squares}
\label{optic} 
\end{figure*}

In Fig.~\ref{optic} we show the contours of the broad-band X-ray
emission of the inner part of the PSPC superimposed onto a digitized deep
optical image from the 3.9-m Anglo Australian Telescope, reproduced by
courtesy of the Anglo Australian Observatory.
A number of point-like sources distributed
throughout the plane of M83 and complex extended emission centered on,
and filling almost the whole optically visible area of the galaxy, are 
evident. 

\begin{figure*}
\mbox{\epsfxsize=80mm\epsfbox{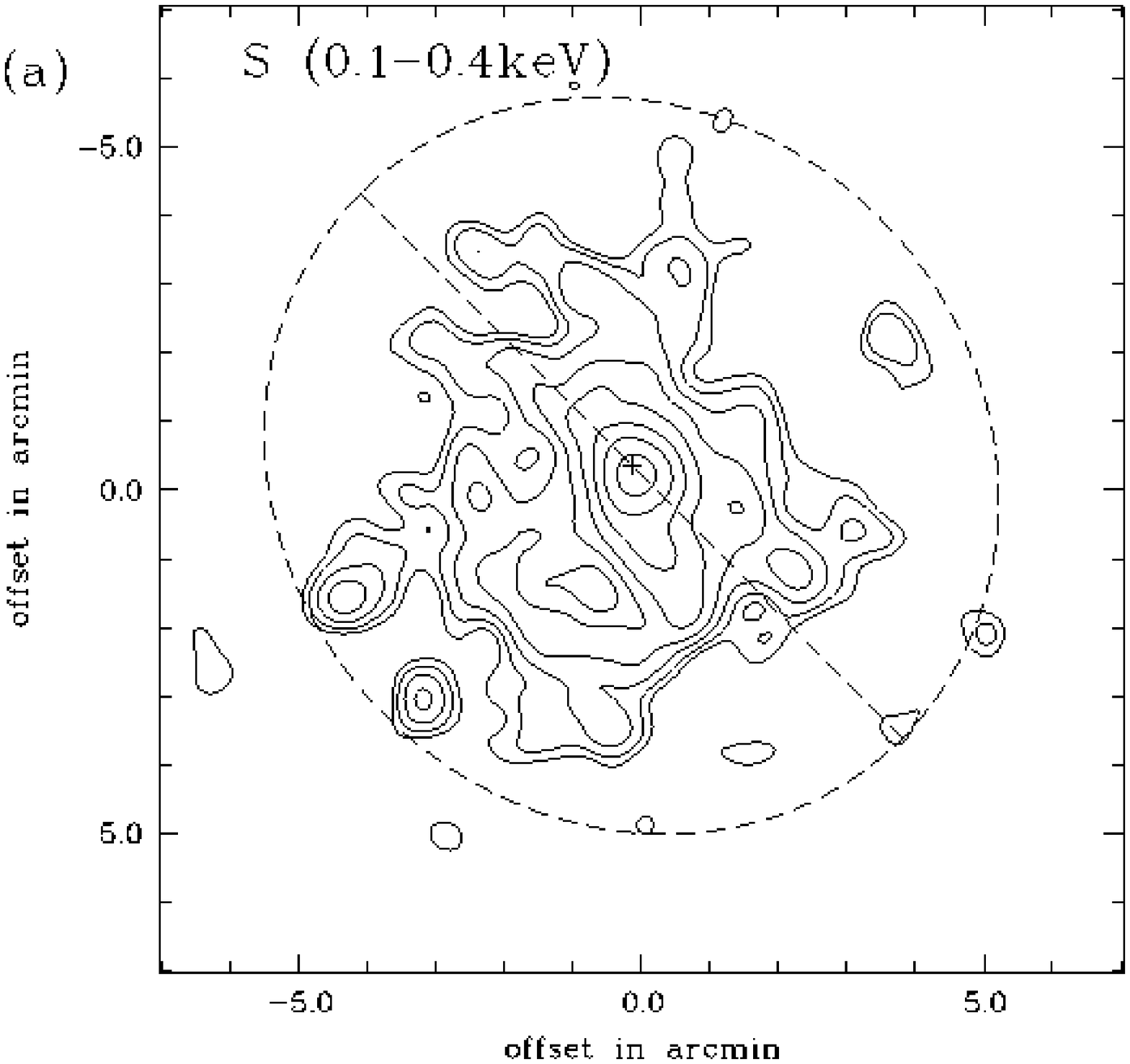}}
\mbox{\epsfxsize=80mm\epsfbox{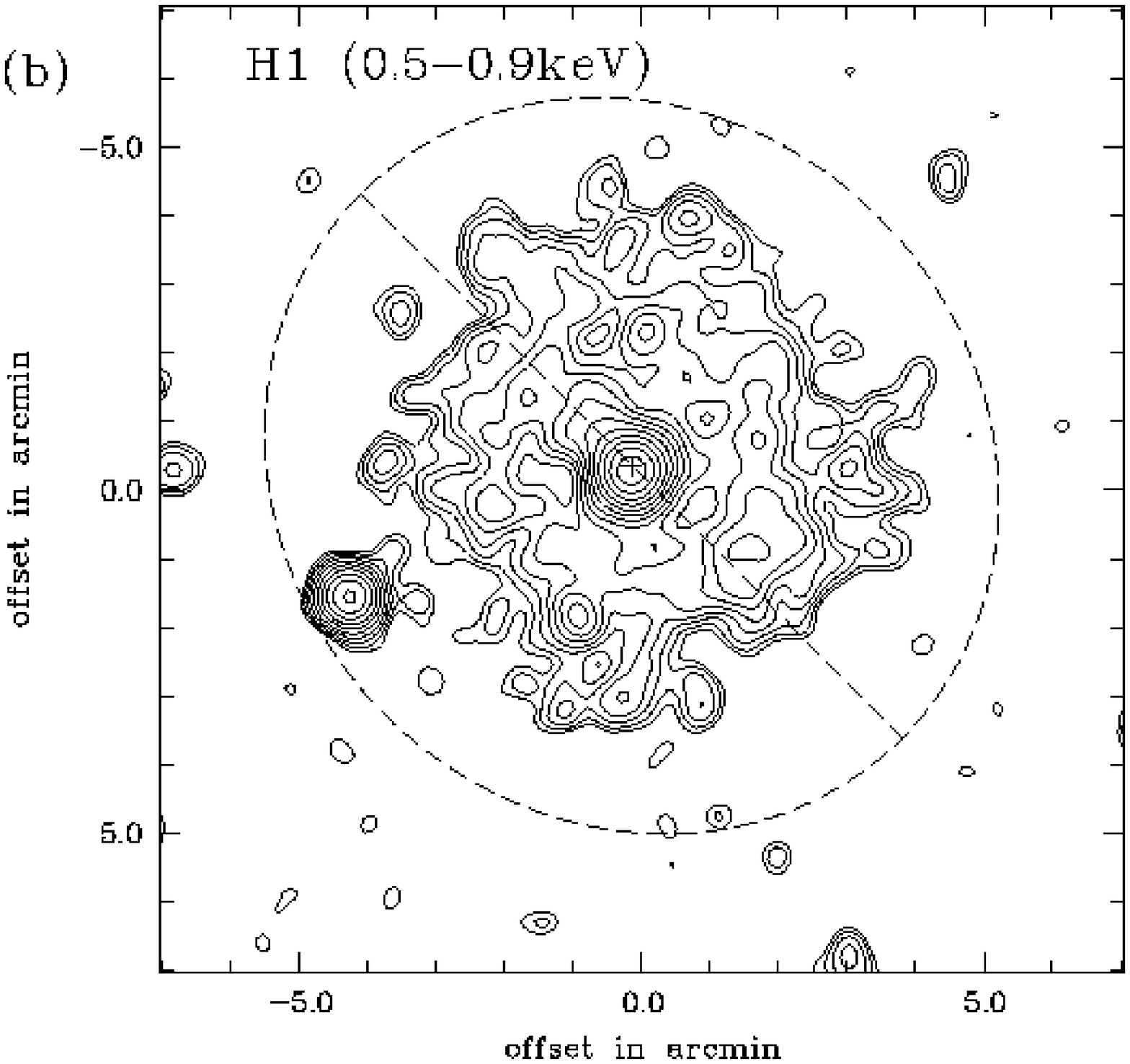}}\vspace*{3mm}
\mbox{\epsfxsize=80mm\epsfbox{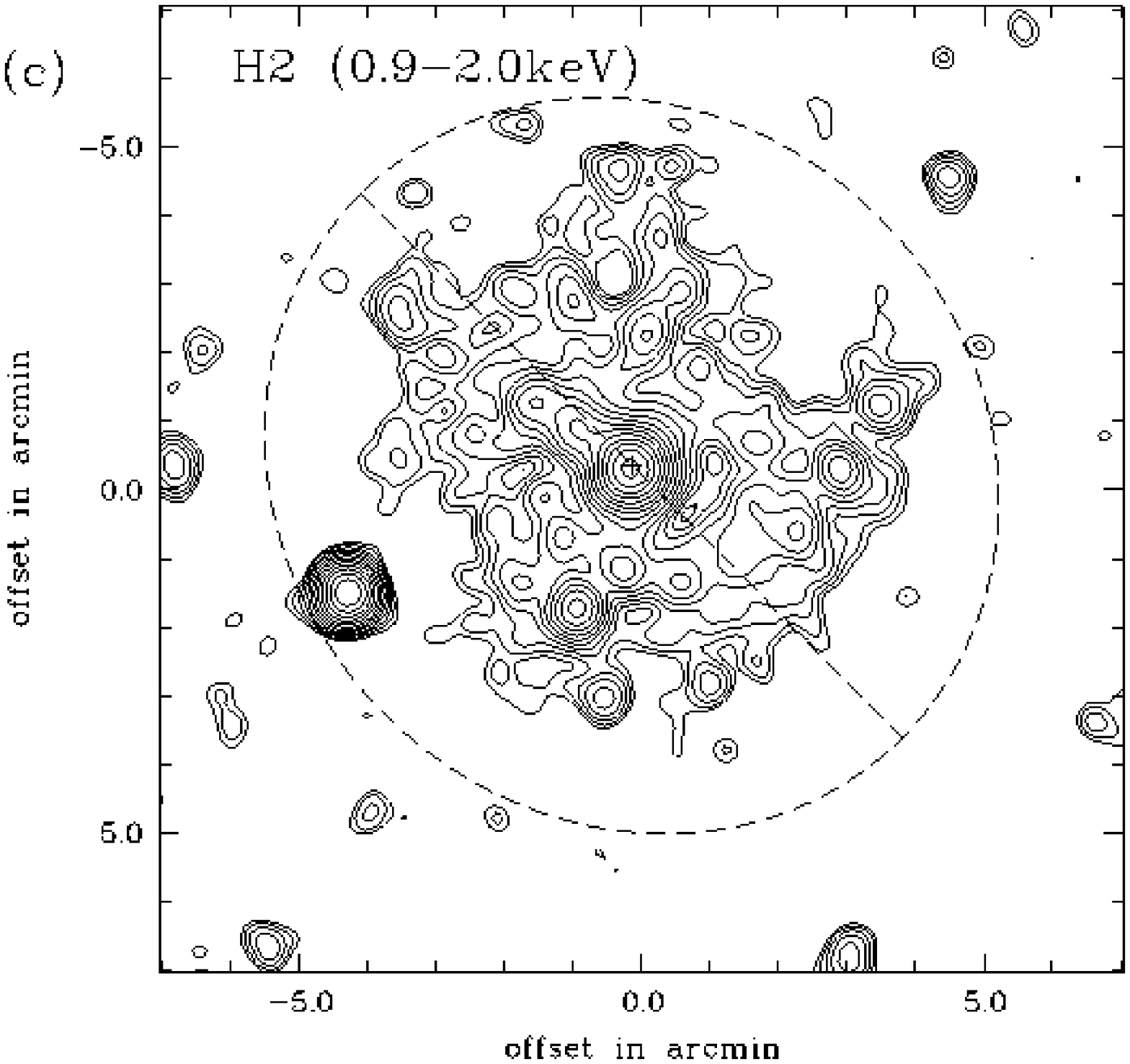}}
\caption[]{Contour plots of the emission region of M83 for the S (a),
H1 (b) and H2 (c) \ROSAT PSPC bands. The images have been smoothed
and exposure-corrected for the individual bands. Contours 
are $(1,1.4,2,2.8,4,...)\times5\cdot10^{-4}$ cts s$^{-1}$ arcmin$^{-2}$
above the background. FWHM and background values are given 
in Table~\ref{baender}.
The dashed ellipse shows the optical extent of the galaxy ($D_{25}$),
and the dashed line the position of the major axis. The position of
the central \ROSAT Source 6 is marked by a cross}
\label{contours}
\end{figure*}

Figure~\ref{contours} gives the contour plots for the S (a), H1 (b) and
H2 (c) \ROSAT energy bands (see Table~\ref{baender}) to study the energy 
dependence of the X-ray emission in more detail.

\subsection{Source detection and identification}
\label{sources}
Source detection and position determination have been performed in the
inner 40\arcmin~ region of the \ROSAT PSPC detector field of view where
the FWHM of the point response function of the X-ray telescope/PSPC
system is smaller than 100\arcsec.
For better spectral sensitivity the source detections have been carried 
out separately in the five \ROSAT energy bands. Details of these EXSAS
detection procedures are given in Pietsch et al. (1994).

We detected a total of 47 sources with a
likelihood of existence $\gid 10$ (see Table~\ref{master} and 
Figs.~\ref{field} and \ref{optic}). The likelihood is defined as 
($-\ln{(1-P)}$) where $P$ is the probability that the observed
distribution of photons originates from a spurious background
fluctuation which in turn is calculated from Poisson statistics.
The source counts and its 1-$\sigma$ errors are determined by the maximum 
likelihood method (cp. Zimmermann et al. 1994). They are 
background subtracted (using the smoothed background
image created by cutting out sources with a cut-out radius of 1.5 FWHM at 
that position and performing a bicubic spline fit to the so-called
cheesed image) but not corrected for mirror efficiency (vignetting).

Only those 13 sources inside of M83 which were 
not flagged as extended by the detection algorithms were accepted and
displayed in Fig.~\ref{optic}. The detection of
point sources inside the diffuse X-ray emission of M83 is difficult 
and the likelihood values of these sources might be underestimated.
It is worth mentioning that the soft point source at RA $\sim13^{\rm
h}37^{\rm m}15^{\rm s}$, Dec $\sim-29\degr55\farcm4$ (J2000) is 
detected with a likelihood value of 7 only, although it is clearly visible in
Fig.~\ref{contours} (a).

Table~\ref{master} gives the parameters of all {\it ROSAT}-detected and
accepted X-ray sources. The position errors (Col. 5) of the sources have 
been calculated in the energy band with the highest likelihood at the 90\%
confidence radius. The likelihood obtained for the sources are
given in Col. 6 together with the corresponding energy band 
in Col. 7. The counts from the sources in that band are listed in Col. 8.
Count rates in the broad band, now corrected for exposure and 
vignetting, are given in Col. 9.

\begin{table*}
\caption[]{X-ray parameters of sources detected in 40\arcmin x40\arcmin~ field centered on M83}
\label{master}
\begin{flushleft}
\begin{tabular}{llllcccrr}
\noalign{\smallskip}
\hline
\noalign{\smallskip}
$\#$&\ROSAT name&RA(J2000)&Dec(J2000)& Error&Lik.&in&Net counts&Count rates\\
& & & & &of exist. &band&in that band&in broad band\\
&&($^{h~~m~~s}$)&(\degr~~~\arcmin~~~\arcsec)&(\arcsec)& & & &$(10^{-3}$s$^{- 1})$\\ 
(1)&(2)&(3)&(4)&(5)&(6)&(7)&(8)&(9)\\
\noalign{\smallskip} \hline \noalign{\smallskip}
1&RX J133644-2950.9&13:36:44.1&-29:50:59& 9&  14&H2&$18.2\pm\pheins5.3$& 1.5$\pm$0.4\\ 
2&RX J133646-2951.9&13:36:46.5&-29:51:55& 6&  33&H&$54.0\pm\pheins9.4$& 3.0$\pm$0.6\\
3&RX J133649-2952.7&13:36:49.7&-29:52:46&10&  10&H2&$25.7\pm\pheins7.5$& 2.1$\pm$0.6\\ 
4&RX J133656-2948.1&13:36:56.6&-29:48:06&13&  16&H1&$19.0\pm\pheins5.2$& 0.5$\pm$0.4\\ 
5&RX J133659-2949.9&13:36:59.6&-29:49:54& 7&  25&H&$52.0\pm\pheins9.9$& 3.0$\pm$0.6\\ 
6&RX J133700-2951.8&13:37:00.7&-29:51:50& 2& 478&H&$753.3\pm34.8$& 40.1$\pm$1.9\\ 
7&RX J133701-2953.4&13:37:01.4&-29:53:24& 8&  13&H&$38.3\pm\pheins9.5$& 2.5$\pm$0.6\\
8&RX J133704-2951.3&13:37:04.3&-29:51:20& 9&  20&H&$74.0\pm14.2$& 4.0$\pm$0.8\\
9&RX J133704-2953.9&13:37:04.6&-29:53:59& 5&  45&H&$82.0\pm11.9$& 4.5$\pm$0.7\\
10&RX J133705-2952.9&13:37:05.8&-29:52:56&10&  11&H&$31.9\pm\pheins8.5$& 2.0$\pm$0.6\\
11&RX J133707-2950.8&13:37:07.3&-29:50:51& 7&  23&H&$58.3\pm11.1$& 3.2$\pm$0.7\\
12&RX J133716-2949.5&13:37:16.6&-29:49:31&11& 114&S&$186.4\pm14.7$& 2.6$\pm$0.5\\
13&RX J133719-2953.6&13:37:19.9&-29:53:38& 3&1143&H&$355.7\pm19.2$& 21.9$\pm$1.1\\
14&RX J133530-2950.5&13:35:30.5&-29:50:31& 2&4201&B&$1962.7\pm46.0$& 107.6$\pm$2.5\\
15&RX J133543-2950.1&13:35:43.2&-29:50:06&18&  10&H&$17.2\pm\pheins5.4$& 1.3$\pm$0.5\\
16&RX J133548-2956.3&13:35:48.9&-29:56:20&20&  20&H2&$18.2\pm\pheins5.3$& 2.1$\pm$0.5\\ 
17&RX J133611-2943.5&13:36:11.8&-29:43:32&10&  66&H&$54.1\pm\pheins8.3$& 4.5$\pm$0.6\\
18&RX J133614-2956.0&13:36:14.3&-29:56:02& 7&  68&H&$42.5\pm\pheins7.0$& 3.1$\pm$0.5\\
19&RX J133615-2957.8&13:36:15.5&-29:57:50&14&  27&H&$29.5\pm\pheins6.4$& 2.4$\pm$0.5\\
20&RX J133618-3001.3&13:36:18.8&-30:01:22&14&  31&H&$26.1\pm\pheins6.0$& 1.9$\pm$0.5\\
21&RX J133619-2934.6&13:36:19.0&-29:34:37&26&  12&H&$17.3\pm\pheins5.3$& 2.3$\pm$0.5\\
22&RX J133627-2955.5&13:36:27.7&-29:55:34&15&  15&H&$20.3\pm\pheins5.7$& 1.4$\pm$0.4\\
23&RX J133633-3003.4&13:36:33.6&-30:03:29& 7& 116&H&$68.3\pm\pheins8.9$& 4.2$\pm$0.6\\
24&RX J133639-2947.6&13:36:39.4&-29:47:39& 8&  33&H&$27.5\pm\pheins5.8$& 2.0$\pm$0.4\\
25&RX J133641-2959.8&13:36:41.0&-29:59:48&14&  11&H&$13.7\pm\pheins4.5$& 1.1$\pm$0.4\\
26&RX J133645-3003.3&13:36:45.9&-30:03:18&20&  27&H2&$17.0\pm\pheins4.8$& 1.3$\pm$0.5\\
27&RX J133646-2959.0&13:36:46.1&-29:59:02& 7&  51&H&$39.6\pm\pheins7.0$& 2.9$\pm$0.5\\
28&RX J133658-2943.5&13:36:58.9&-29:43:30&10&  24&H2&$16.3\pm\pheins4.4$& 1.8$\pm$0.4\\
29&RX J133711-2935.6&13:37:11.3&-29:35:40&15&  31&H&$45.7\pm\pheins8.4$& 2.3$\pm$0.6\\
30&RX J133715-3002.7&13:37:15.4&-30:02:43&11&  41&H&$31.9\pm\pheins6.5$& 2.0$\pm$0.5\\
31&RX J133725-2958.8&13:37:25.2&-29:58:50&12&  15&H2&$11.8\pm\pheins3.8$& 1.2$\pm$0.4\\
32&RX J133730-2942.5&13:37:30.2&-29:42:31&19&  11&H&$18.1\pm\pheins5.5$& 1.7$\pm$0.5\\
33&RX J133731-2951.8&13:37:31.5&-29:51:52& 7&  44&H&$34.4\pm\pheins6.6$& 2.4$\pm$0.4\\
34&RX J133736-2948.2&13:37:36.7&-29:48:15& 9&  39&H&$36.5\pm\pheins6.9$& 2.3$\pm$0.5\\
35&RX J133738-3003.3&13:37:38.9&-30:03:21&14&  23&H&$26.6\pm\pheins6.3$& 1.6$\pm$0.5\\
36&RX J133739-2943.3&13:37:39.4&-29:43:18&16&  13&H&$18.2\pm\pheins5.3$& 1.2$\pm$0.4\\
37&RX J133742-2951.5&13:37:42.4&-29:51:34& 8&  44&H&$36.0\pm\pheins6.7$& 2.3$\pm$0.5\\
38&RX J133744-3005.7&13:37:44.2&-30:05:47&25&  11&H&$18.3\pm\pheins5.6$& 1.5$\pm$0.5\\
39&RX J133744-2952.9&13:37:44.6&-29:52:59&10&  39&H&$33.5\pm\pheins6.5$& 2.5$\pm$0.5\\
40&RX J133749-2956.6&13:37:49.6&-29:56:38&19&  13&H&$19.8\pm\pheins5.6$& 1.3$\pm$0.4\\
41&RX J133803-2943.9&13:38:03.4&-29:43:59&12&  38&H2&$140.0\pm14.7$& 7.8$\pm$0.9\\
42&RX J133806-2957.6&13:38:06.0&-29:57:38&13&  43&H&$49.9\pm\pheins8.4$& 4.0$\pm$0.7\\
43&RX J133806-2944.3&13:38:06.1&-29:44:22&10& 161&H&$298.2\pm20.4$& 17.8$\pm$1.3\\
44&RX J133806-2943.6&13:38:06.6&-29:43:40& 9&  78&H&$128.7\pm13.4$& 7.1$\pm$0.8\\
45&RX J133806-2945.2&13:38:06.8&-29:45:12&10&  67&H&$103.6\pm12.1$& 6.1$\pm$0.8\\
46&RX J133809-2957.6&13:38:09.2&-29:57:41&19&  10&B&$50.8\pm13.0$& 2.7$\pm$0.7\\
47&RX J133811-2953.9&13:38:11.6&-29:53:59&17&  21&H&$30.3\pm\pheins6.9$& 1.9$\pm$0.5\\
\noalign{\smallskip} \hline
\end{tabular}
\end{flushleft}
\end{table*}

The count-to-energy conversion factor for e.g. a 5 keV thermal bremsstrahlung
spectrum, corrected for Galactic foreground absorption in the direction
towards M83, is $1.75\cdot10^{-11}$ erg cm$^{-2}$ cts$^{-1}$. Under this
assumption a $1\cdot10^{-3}$ cts s$^{-1}$ source at the distance of M83
would have an X-ray luminosity in the broad \ROSAT energy band
of $1.66\cdot10^{38}$ erg s$^{-1}$. 

Outside the 40\arcmin~ detection area two more strong X-ray sources are
visible at RA $\sim13^{\rm h}38^{\rm m}43^{\rm s}$,
Dec $\sim-29\degr33\farcm5$ (partly visible in the upper left corner of
Fig.~\ref{field}) and RA $\sim13^{\rm h}34^{\rm m}58^{\rm s}$,
Dec $\sim-29\degr55\farcm3$ (J2000). Due to the large off-axis angles
these positions have large errors ($\sim2\arcmin$). The first of
these sources can be identified as the F5V star HR\,5128 (see above) which
was already detected as the \EINSTEIN Source I8. The second one corresponds to
\EINSTEIN Source I9 (Trinchieri et al. 1985) where it was interpreted
as probably a 9th magnitude star.

The starburst nucleus of M83 was detected as an extended X-ray Source 6 and 
is the strongest source in that galaxy. Its X-ray position coincides very
well with the central radio continuum source at RA $=13^{\rm h}37^{\rm
m}00\fs25$, Dec $=-29\degr51\arcmin51\farcs3$ (J2000) (Condon et al.
1982). 

We searched for X-ray emission from cataloged historical supernovae in
M83 (Richter \& Rosa 1984, and references therein): 
no enhanced X-ray emission at the positions of the radio detections of 
SN1923a, SN1950b, SN1957d, SN1968l and SN1983n was found in
any of the \ROSAT energy band images. 

We also compared the {\it ROSAT}-detected sources with earlier \EINSTEIN
observations, maps of H$\alpha$, \HI emission (Tilanus \& Allen 1993) 
and optical ESO-SRC plates (Table~\ref{ident}).
\begin{table*}
\caption[]{Point Source Identification in and around M83}
\label{ident}
\begin{flushleft}
\begin{tabular}{lrl}
\noalign{\smallskip} \hline \noalign{\smallskip}
\multicolumn{3}{c}{Sources in M83}\\
\noalign{\smallskip} \hline \noalign{\smallskip}
Source & Luminosity$^a$            & Environment and Comments\\
\noalign{\smallskip} \hline \noalign{\smallskip}
      1& $2.5\pm0.6$& \HI hole, at the edge of an \HII region\\
      3& $3.5\pm1.0$& at the edge of an \HII region\\
      4& $0.8\pm0.7$& at the edge of an \HII region\\
      5& $5.0\pm1.0$& near \HI hole\\
      6& $66.6\pm3.2$& nucleus, \HI in absorption, \HII region, T: \EINSTEIN Source H1 ($\Delta$RA=2\arcsec, $\Delta$Dec=6\arcsec),\\
       &      & $L_{\rm x}$(0.5-3.0\,keV)$=113\cdot10^{38}$ erg s$^{-1}$, C: radio position (6\arcsec, 1\arcsec)\\
      7& $4.2\pm1.0$& \HI hole\\
      8& $6.6\pm1.3$& T: \EINSTEIN Source H4 (0\arcsec, -2\arcsec), $L_{\rm x}$(0.5-3.0\,keV)$=13\cdot10^{38}$ erg s$^{-1}$\\
      9& $7.5\pm1.2$& \HII region, CRB: near therm. radio Source 8 (19\arcsec, 3\arcsec)\\
     11& $5.3\pm1.2$& \HII region\\
     13& $36.4\pm1.8$& \HI in absorption, T: \EINSTEIN Sources I2 (10\arcsec, -2\arcsec) $L_{\rm x}$(0.5-3.0\,keV)$=39\cdot10^{38}$ erg s$^{-1}$\\
       &      & or H2 (3\arcsec, 8\arcsec) $L_{\rm x}$(0.5-3.0\,keV)$=24\cdot10^{38}$ erg s$^{-1}$, ESO-SRC plates: diffuse faint object\\
\noalign{\smallskip} \hline \noalign{\smallskip}
\multicolumn{3}{c}{Sources around M83}\\
\noalign{\smallskip} \hline \noalign{\smallskip}
Source & Flux$^a$             & Environment and Comments \\
\noalign{\smallskip} \hline \noalign{\smallskip}
     14& $188.3\pm4.4$& T: \EINSTEIN Source I6 (8\arcsec, 4\arcsec) $f_{\rm x}$(0.5-3.0\,keV)$=3.2\cdot10^{-12}$ erg cm$^{-2}$ s$^{-1}$, G: AGN MS1332.6-2935\\
     16& $3.7\pm0.9$& ESO-SRC plates: several objects\\
     23& $7.4\pm1.1$& ESO-SRC plates: diffuse faint object\\
     27& $5.1\pm0.9$& ESO-SRC plates: diffuse faint object\\
     28& $3.2\pm0.7$& ESO-SRC plates: faint object\\
     41,43-45& $67.9\pm3.3$& T: \EINSTEIN Source I7 (background cluster, z=0.189, H$_{\circ}=50$~km/s Mpc)\\
       &      & $L_{\rm x}$(0.5-3.0\,keV)$=7.0\cdot10^{43}$ erg
       s$^{-1}$, $L_{\rm x}$(0.1-2.4\,keV)$=(19.4\pm0.5)\cdot10^{43}$ erg
       s$^{-1}$\\
       &      & ESO-SRC R-plate: $\ga$3 diffuse objects (nothing visible on J-plate)\\
     42,46& $11.7\pm1.8$& ESO-SRC plates: two faint objects\\
\noalign{\smallskip} \hline
\end{tabular}
\end{flushleft}
$^a$ - Assuming kT=5\,keV, $N_{\rm H}=3.97\cdot10^{20}$\,cm$^{-2}$,
distance to M83=8.9\,Mpc. Luminosities in units of $10^{38}$ erg s$^{-1}$, 
fluxes in $10^{-14}$ erg cm$^{-2}$ s$^{-1}$\\
$^b$ - T: Trinchieri et al. 1985; C: Condon et al. 1982; CRB: Cowan et al. 1994; G: Gioia et al. 1990
\end{table*}

\subsection{Spatial analysis of the extended emission}
\label{extended}
Besides the detected X-ray sources, M83 shows complex extended X-ray
emission in all \ROSAT energy bands (see Fig.~\ref{contours}) covering
essentially the whole optically visible galaxy. 

The S-band contours are rather smooth and (neglecting Source 13 and the
source at RA $\sim13^{\rm h}37^{\rm m}15^{\rm s}$, Dec 
$\sim-29\degr55\farcm4$, see Fig.~\ref{optic}) probably do not contain 
a significant contribution from point sources because their soft X-ray 
emission is absorbed within the galactic disk of M83. 
In the H1- and H2-band the point sources are more clearly visible.
This is partly caused by the harder spectra of the point sources but
also because of the higher spatial resolution of the PSPC at these 
energies.
\subsubsection{Comparison with \EINSTEIN HRI observations}
\label{comp_einstein}
The soft emission shows a central ridge structure starting from the
nucleus in directions P.A.$\sim40\degr$ as well as P.A.$\sim195\degr$ 
(Fig.~\ref{contours}a).
In contrast, the hard emission from this inner part of the galaxy is
strongly asymmetric with respect to the central nuclear source: the
north-eastern part of the ridge structure is still visible, but the
south-western protrusion disappears. Here in the H1 band we see a reduction
in the emission that becomes even more visible in the H2 band. In
addition, the northern spur observed in the H2 band is pointing to a 
slightly different direction (P.A.$\sim55\degr$, Fig.~\ref{contours}b, c). 

\begin{figure}
{\centering\psfig{figure=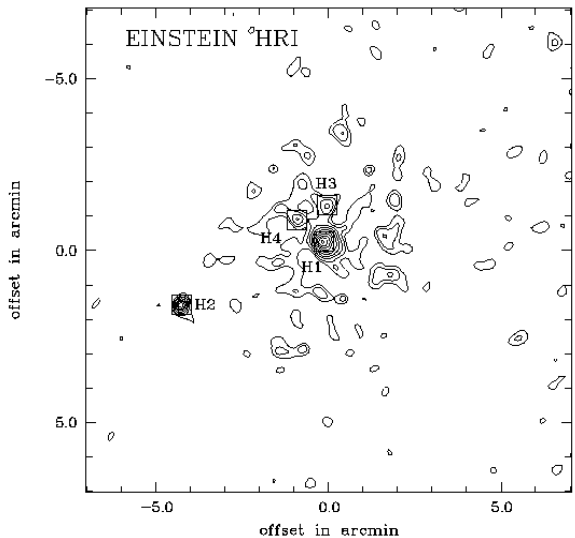,width=85mm}}
\caption[]{\EINSTEIN HRI contours of M83, adaptively filtered with Gaussian
functions of FWHM$\le8\farcs6$ and aligned with our \ROSAT observations. 
Contours of $(1,1.4,2,2.8,4,...)\times3\sigma$ above the background are 
plotted (cp Fig.~4 in Trinchieri et al. (1985)). Positions of the 
\EINSTEIN-HRI-detected sources are marked as squares}
\label{einstein}
\end{figure}

To study the central region at higher angular resolution, we reanalysed 
\EINSTEIN HRI observations (using data obtained through the High Energy
Astrophysics Science Archive Research Center Online Service, provided by
the NASA Goddard Space Flight Center) which were already published by 
Trinchieri et al. (1985). We combined the two \EINSTEIN observations 
(1980 Jan 15-16: 24575~s \& 1981 Feb 13: 20003~s) after
correcting for a positional offset, aligned them with our \ROSAT maps and
used adaptive filtering (cf. Ehle et al. 1995) to produce a new map 
which is displayed in Fig.~\ref{einstein} as a contour plot.

In the \EINSTEIN image we identify the north-eastern spur and
the two emission peaks that are visible in the H2 band emission 
(Fig.~\ref{contours}c). It is worth mentioning that we do not
detect \EINSTEIN Source H3 in any of our \ROSAT energy bands and 
therefore conclude that this source is variable. 
The \EINSTEIN HRI contour plot shows
emission extending from the nucleus in a south-western direction
(P.A.$\sim208\degr$), similar to the ridge structure in the soft
image, but its radial extent is much smaller, probably owing to
lack of sensitivity. Slightly enhanced X-ray emission south of the 
nuclear region is visible in the H1 band and is probably associated 
with the south-western spur that is visible in the \EINSTEIN 
observations as well. 
\subsubsection{Radial and azimuthal distribution of the diffuse X-ray emission}
\label{distribution}
To quantify the distribution of the diffuse X-ray emission in our \ROSAT maps, 
we subtracted the detected point-like sources and integrated the remaining 
diffuse emission separately in the north-western and south-eastern half of 
M83 out to a galactic radius of 5'. The two halves were separated by the 
galaxy's major axis.
While the emission in the two hard bands is enhanced in the 
north-western part by about $3\%$ relative to the south-eastern part, 
the soft-band emission shows the opposite asymmetry, with a lack of 
soft emission in the north-western half of M83 (decreased by about $3\%$).

\begin{figure}
{\centering\psfig{figure=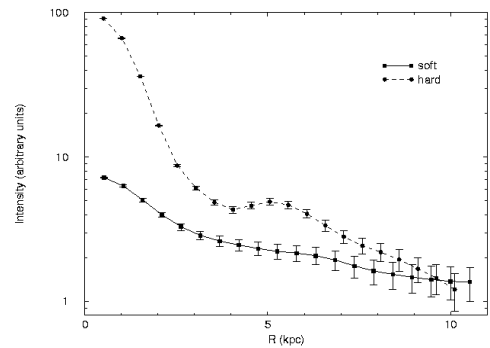,width=85mm,angle=-90}}
\caption[]{Radial profiles of the extended component of the X-ray
emission of M83. The plot gives the intensities (in arbitrary units)
in circular annuli around the central source for the soft and hard
energy range, separately}
\label{profile}
\end{figure}

In order to investigate if the diffuse X-ray emission of M83 originates from
sources that are also responsible for the production of cosmic rays (as
was suggested by Fabbiano et al. (1982) in the case of peculiar
galaxies) we compared the extent of the radio continuum (Neininger 1992)
and the diffuse X-ray emission. 
Figure~\ref{profile} gives the the radial profiles of the extended X-ray 
emission for the soft and hard energy range, calculated by
averaging counts in rings of 12\arcsec\ (0.5 kpc) which are centered on
the position of the central X-ray source. Whereas the soft emission is smoothly
decreasing with radius, the profile of the hard emission component has
a local minimum at $\sim4$~kpc (corresponding to the reduction in the
emission south-west of the nucleus (cp. Fig.~\ref{contours}c)) followed
by an enhancement. This local maximum is also visible as
a distinct `hump' in the radial distribution of the optical surface brightness
(Talbot et al. 1979) where this region was found to contain the bluest light
in the galaxy outside the nucleus. The radio continuum profiles 
at $\lambda2.8$~cm (Neininger 1992), however, are very smooth over the 
whole radial extent of M83.
 
The scale length of the extended hard-band X-ray emission outside a
radius of 5.5~kpc is $l_{\rm x}=3.4\pm0.3$~kpc (neglecting the nuclear
region with $l_{\rm x}= 1.1\pm0.2$~kpc as well as circular annuli
around the local minimum at $\sim4$ kpc). The scale length of the
$\lambda2.8$~cm radio emission is $l_{\rm syn}\sim4.4$~kpc.
Neininger (1992) separated the thermal and nonthermal radio
continuum emission of M83 and calculated for the thermal emission at
$\lambda2.8$~cm a scale length of $l_{\rm th}\sim3.2$~kpc, consistent
with that of the hard-band diffuse X-ray emission.

An important finding is that the radial profile of the soft diffuse
X-ray emission shows a much larger scale length that is best
fitted with a Gaussian of half FWHM $\sim10-15$~kpc.
In addition, the `hump' in the profile of the hard X-ray emission does
not exist in the soft energy image.
\subsubsection{Separation of emission components}
\label{compo}
We calculated the count rates and hardness ratios (defined as hardness 
ratio 1 HR1=(hard-soft)/(hard+soft) and hardness ratio 2 
HR2=(hard2-hard1)/(hard2+hard1), see Table~\ref{hr}) for the 
diffuse emission of M83. The extended emission in the different energy bands
was integrated over a ring with an inner radius of 1\arcmin~ and
outer radius of 5\arcmin~ and corrected for the background and time
of exposure. The same was done for the nuclear region with radius 1\arcmin~
separately. The definition of the extent of the nuclear region is somewhat
arbitrary, and it results from the qualitative inspection of the contour
maps. 
We should note that this and all following calculations making use of the
so called `diffuse emission' are based on upper limits of this emission
component. Part of it may be due to sources below our detection threshold
(cf. Sect.~\ref{discu_point}).
  
To determine the emission resulting from detected point-like sources,
we added up the count rates in the different energy bands as they were
obtained from our source detection (Table~\ref{hr}).
\begin{table*}
\caption[]{Components of the X-ray emission from M83}
\label{hr}
\begin{flushleft}
\begin{tabular}{llcc}
\hline
\noalign{\smallskip}
Emission component&Count rates  &\multicolumn{2}{c}{Hardness ratios}\\
		  & $10^{-3}$ cts s$^{-1}$ & HR1 & ~HR2\\
\hline
\noalign{\smallskip}
Nuclear region    & $103\pm2$      & $0.76\pm0.02$ & $\phdoteins0.03\pm0.02$\\
Extended emission & $197\pm4$      & $0.49\pm0.03$ & $-0.10\pm0.02$\\
Point-like sources&$\pheins29\pm2$ & $0.34\pm0.05$ & $\phdoteins0.28\pm0.07$\\
Total emission    & $329\pm5$      & $0.56\pm0.02$ & $-0.02\pm0.02$\\
\hline
\noalign{\smallskip}
\end{tabular}
\end{flushleft}
\end{table*}
The total X-ray emission of M83 is dominated by the complex extended 
emission in all \ROSAT energy bands. All
detected sources (excluding the nuclear Source 6 and Source 13) contribute
only 9\% to the total count rate. The hardness ratios show that the
point-like sources in general have hard spectra and that the nuclear
emission is harder than the extended emission. 

\subsection{Spectral analysis: derivation of X-ray colors and fits to models}
\label{ana-spectral}
Nearly all sources (1, 2, 4-11) in M83 are most likely bright X-ray 
binaries showing strong absorption of the soft-band flux
as expected from their position in the galactic disk of M83. 
Source 12 shows enhanced soft X-ray emission. The very strong
X-ray source 13 might be an AGN, a quasar, or a radio galaxy.
Only this source and the nucleus are strong
enough to allow a detailed spectral analysis. The best fit to
Source 13 is a thermal plasma spectrum with $N_{\rm H}$ fixed to the
Galactic foreground value, yielding a temperature of $(3.1\pm^{2.4}_{0.6})
\cdot10^7$ K and a flux density of $(3.5\pm1.4)\cdot10^{-13}$ erg cm$^{-2}$
s$^{-1}$ (errors correspond to the 68.3\% confidence level of our best fit
with $\chi^2/\nu=0.68$ and 14 degrees of freedom).

A more detailed spectral analysis of other single point sources 
is very difficult because of the unknown contribution of the underlying 
diffuse emission. 

To study the spectra of the diffuse extended X-ray emission we excluded 
circular regions (with a radius 1.5$\times$ FWHM of the point response 
function) at the positions found by our source detection.
Source 13 was excluded because it is most likely a background source.
We also excluded the central part of M83 (radius $1\arcmin\cor2.6$ kpc
around the position of the central Source 6) which we shall discuss 
as the nuclear region in Sect.~\ref{res_nuc}. The remaining emission, 
integrated out to a galactic radius of $5\arcmin\cor12.9$ kpc is discussed as
extended emission in Sect.~\ref{res_diff}. Background values were 
derived by integrating counts in nearby emission-free regions.
The resulting count rate spectra were binned so that each energy 
interval had a signal-to-noise ratio larger than 5.

\subsubsection{Nuclear region}
\label{res_nuc}
\begin{figure}
{\centering\psfig{figure=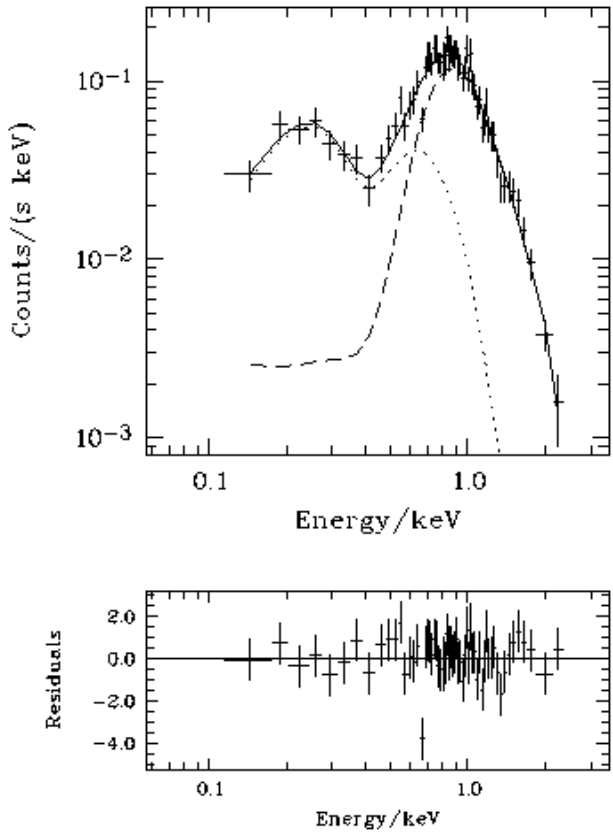,width=85mm}}
\caption[]{Observed \ROSAT PSPC spectrum of the nuclear area of M83
between 0.1 and 2.4~keV (the 2200 photons are divided in 63 bins)
together with the two-temperature fit (see Table~\ref{spektrum}).
The contribution of the low-temperature component is indicated separately
as the dotted and that of the high-temperature component as the dashed
curve. The lower part of the diagram shows residuals in units of standard
deviations}
\label{specnuc}
\end{figure}

\begin{figure}
{\centering\psfig{figure=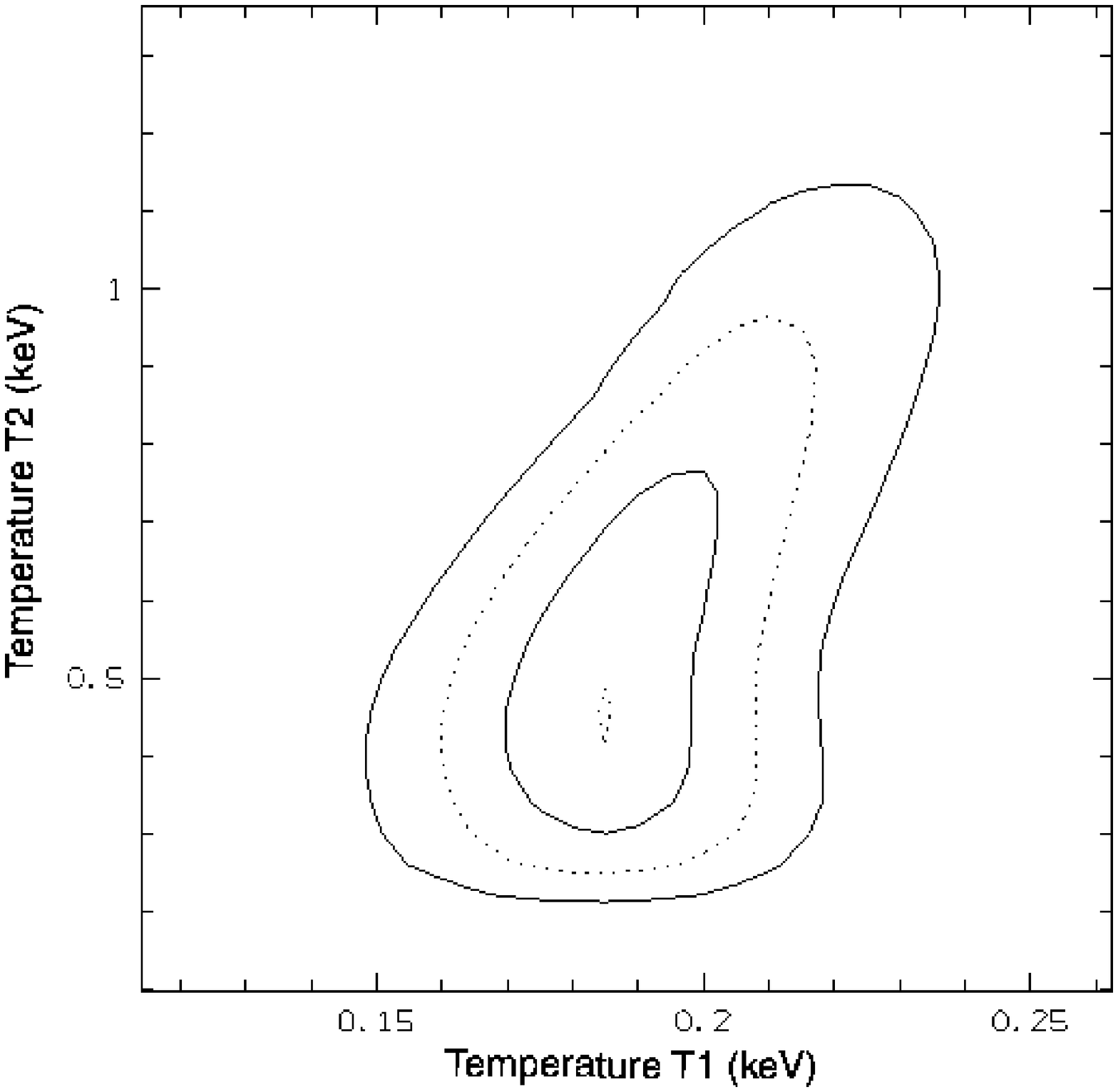,width=85mm,angle=-90}}
{\centering\psfig{figure=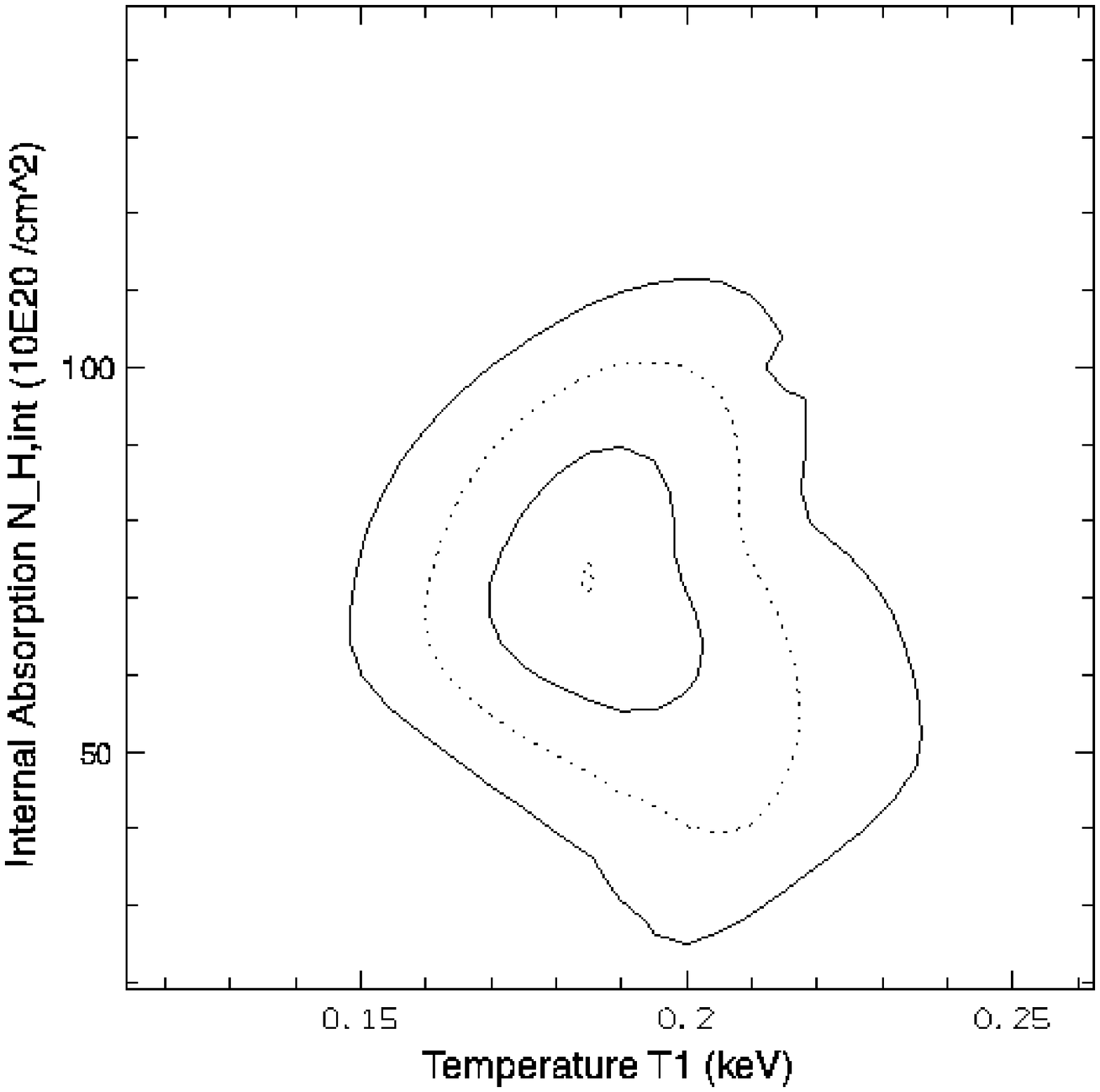,width=85mm,angle=-90}}
\caption[]{$\chi^2$ contours for the model fit (a combination of a thermal
Raymond-Smith plasma of temperature $T_1$ with an internally absorbed
thermal plasma of temperature $T_2$, both affected by Galactic foreground
absorption) to the spectrum of the nuclear area of M83. The region of the
$\chi^2$ minimum is indicated by the inner dashed contour followed by
contours corresponding to 68.3\%, 96.5\% and 99.7\% confidence levels.
The number of ``interesting parameters" (cf. Avni 1976) is three. The 
{\it{top}} panel shows the plot of the parameter set $T_1$ and $T_2$, 
the {\it{bottom}} that of $T_1$ and $N_{\rm H}^{\rm int}$}
\label{nucconf}
\end{figure}

The background subtracted energy spectrum of the photons 
originating from the nuclear area cannot be fitted with sufficient
statistical reliability by simple spectral models, like a power-law, 
a thermal bremsstrahlung or a thin thermal Raymond-Smith model with `cosmic'
abundances (Raymond \& Smith 1977). Details of the used models are
given by Zimmermann et al. (1994). Table~\ref{spektrum} lists the fit results. 
Furthermore, for the thermal plasma model the best-fit foreground 
absorption is too low compared to the Galactic foreground value. 
However, an acceptable fit is produced by the combination of a thermal
Raymond-Smith plasma of temperature $T_1$ with an internally absorbed 
thermal plasma of temperature $T_2$, both affected by Galactic foreground 
absorption. Again, the model assumed `cosmic' abundances to reduce the 
number of free parameters (see Table~\ref{spektrum} for best-fit parameters 
and Fig.~\ref{specnuc}). As the largest uncertainties of the fit come both 
from the dependence of $T_1$ and $T_2$ as well as from that between
$T_1$ and the internal absorption ($N_{\rm H}^{\rm int}$), Fig.~\ref{nucconf} 
shows confidence contours for these two sets of parameters. 
 
The physical motivation for a two-component thermal plasma model is the 
expectation that one
observes a superposition of a soft halo and hard disk emission in face-on
galaxies, as is supported by studies of edge-on galaxies like NGC\,4631
(Wang et al. 1995, Vogler \& Pietsch 1996) 
and NGC\,253 (Pietsch 1993) where the differentiation
of disk and halo regions is possible. Our best-fit model
explains the X-ray emission originating from a warm gas component
($0.18\pm^{0.02}_{0.01}$~keV$\cor2.1\cdot10^6$~K; probably located in the 
halo region) and hot plasma ($0.45\pm^{0.24}_{0.13}$~keV$\cor5.2\cdot10^6$~K; 
probably the hot gas near
to the nucleus and in the disk of M83). The harder X-ray emission from
the hot gas component is strongly internally absorbed. 
This agrees with the result that the nuclear region shows a significantly 
higher HR1 than the point-like sources and the extended emission.
Using the best-fit parameters,
the X-ray luminosity of the M83 nuclear region in the broad \ROSAT
band is $L_{\rm x}= 1.6\cdot10^{40}$ erg s$^{-1}$ (53\% due to
the cooler and 47\% due to the hotter gas component).
The luminosity of the diffuse soft component is
$L_{\rm x}$(soft)$= 0.4\cdot10^{40}$ erg s$^{-1}$ and originates almost
only from the lower-temperature plasma. The soft X-ray luminosity is used
to estimate the physical parameters of the emitting gas in
Sect.~\ref{discu_extended}.

\subsubsection{Diffuse soft emission due to hot gas}
\label{res_diff}
We assume that, after subtraction of the detected sources and the nuclear
area, the remaining broad-band emission of M83 is a mixture of hot thermal
gas (in the disk and in the halo) and weak point sources that might not
be detected in the present observations. Trying to fit the observed photon
spectra with simple models again gave no acceptable results. However,
the combined two-temperature spectral model already discussed for
the nuclear region yields a good description of the
observed spectra. The best-fit parameters for the extended emission are
given in Table~\ref{spektrum}.
 
\begin{table*}
\caption[]{Spectral parameters of the nuclear area and extended emission
of M83. Errors are corresponding to the 68.3\% confidence levels}
\label{spektrum}
\begin{flushleft}
\begin{tabular}{lllllllll}
\hline
\noalign{\smallskip}
Model &\multicolumn{2}{c}{$N_{\rm H}^{(a)}$} & Index &\multicolumn{2}{c}{$kT^{(b)}$}& $\nu^{(c)}$ & $\chi^2/\nu$ & $L_{\rm x}({\rm soft})^{(d)}$\\
& external & internal &  & $kT_1$ & $kT_2$ &  &  \\
\noalign{\smallskip}
\hline
\noalign{\smallskip}
\multicolumn{9}{c}{Nuclear Area}\\
Thermal bremsstrahlung     &$6.8\pm^{0.5}_{0.4}$&     &      &$0.60\pm^{0.04}_{0.04}$&     & 60 & 2.1 &       \\
Power-law                  &$9.6\pm^{0.5}_{0.4}$&     &$3.0\pm^{0.1}_{0.1}$&     &     & 60 & 2.7 &       \\
Raymond-Smith              &$1.1\pm^{0.2}_{0.2}$&     &      &$0.94\pm^{0.07}_{0.03}$&     & 60 & 1.7 &       \\
Raymond-Smith$^{(e)}$          & 3.97&     &      &$0.93\pm^{0.09}_{0.02}$&     & 61 & 4.3 &       \\
Two-temp. Raymond-Smith$^{(e)}$& 3.97&$72.7\pm^{6.6}_{7.8}$&      &$0.18\pm^{0.02}_{0.01}$&$0.45\pm^{0.24}_{0.13}$& 58 & 0.9 & 0.4  \\
\noalign{\smallskip}
\hline
\noalign{\smallskip}
\multicolumn{9}{c}{Extended Emission}\\
Thermal bremsstrahlung     &$5.3\pm^{0.3}_{0.4}$&     &      &$0.44\pm^{0.03}_{0.02}$&     & 58 & 1.2 &       \\
Power-law                  &$8.7\pm^{0.6}_{0.7}$&     &$3.6\pm^{0.2}_{0.1}$&     &     & 58 & 1.6 &       \\
Raymond-Smith              &$1.0\pm^{0.4}_{0.4}$&     &      &$0.37\pm^{0.05}_{0.02}$&     & 58 & 4.1 &       \\
Raymond-Smith$^{(e)}$          & 3.97&     &      &$0.30\pm^{0.03}_{0.01}$&     & 59 & 5.5 &       \\
Two-temp. Raymond-Smith$^{(e)}$& 3.97&$75.4\pm^{10.6}_{\pheins9.4}$&      &$0.18\pm^{0.02}_{0.01}$&$0.49\pm^{0.23}_{0.16}$& 56 & 0.9 & 1.4  \\
\noalign{\smallskip}
\hline
\end{tabular}
\end{flushleft}
$^{(a)}$ in units of $10^{20}$ cm$^{-2}$\\
$^{(b)}$ in units of keV\\
$^{(c)}$ numbers of degrees of freedom\\
$^{(d)}$ luminosity in the soft (0.1-0.4 keV) energy range (for a M83 distance of
8.9 Mpc), corrected for Milky Way absorption, in units of
$10^{40}$ erg s$^{-1}$\\
$^{(e)}$ external $N_{\rm H}$ fixed to Galactic value\\
\end{table*}
\begin{figure}
{\centering\psfig{figure=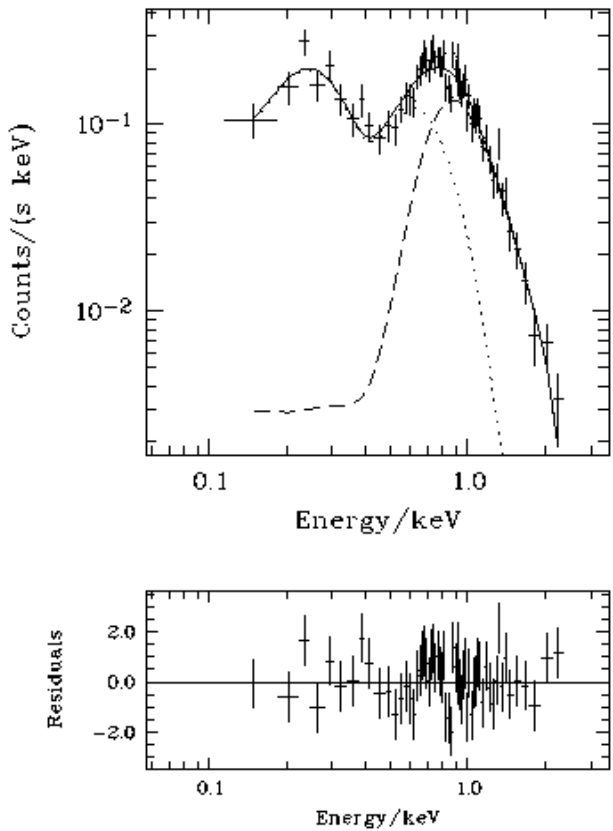,width=85mm}}
\caption[]{Observed \ROSAT PSPC spectrum of the extended emission of M83
between 0.1 and 2.4~keV (the 5600 photons are divided in 61 bins)
together with the two-temperature fit (cp. Fig.~\ref{specnuc})}
\label{specrest}
\end{figure}
\begin{figure}
{\centering\psfig{figure=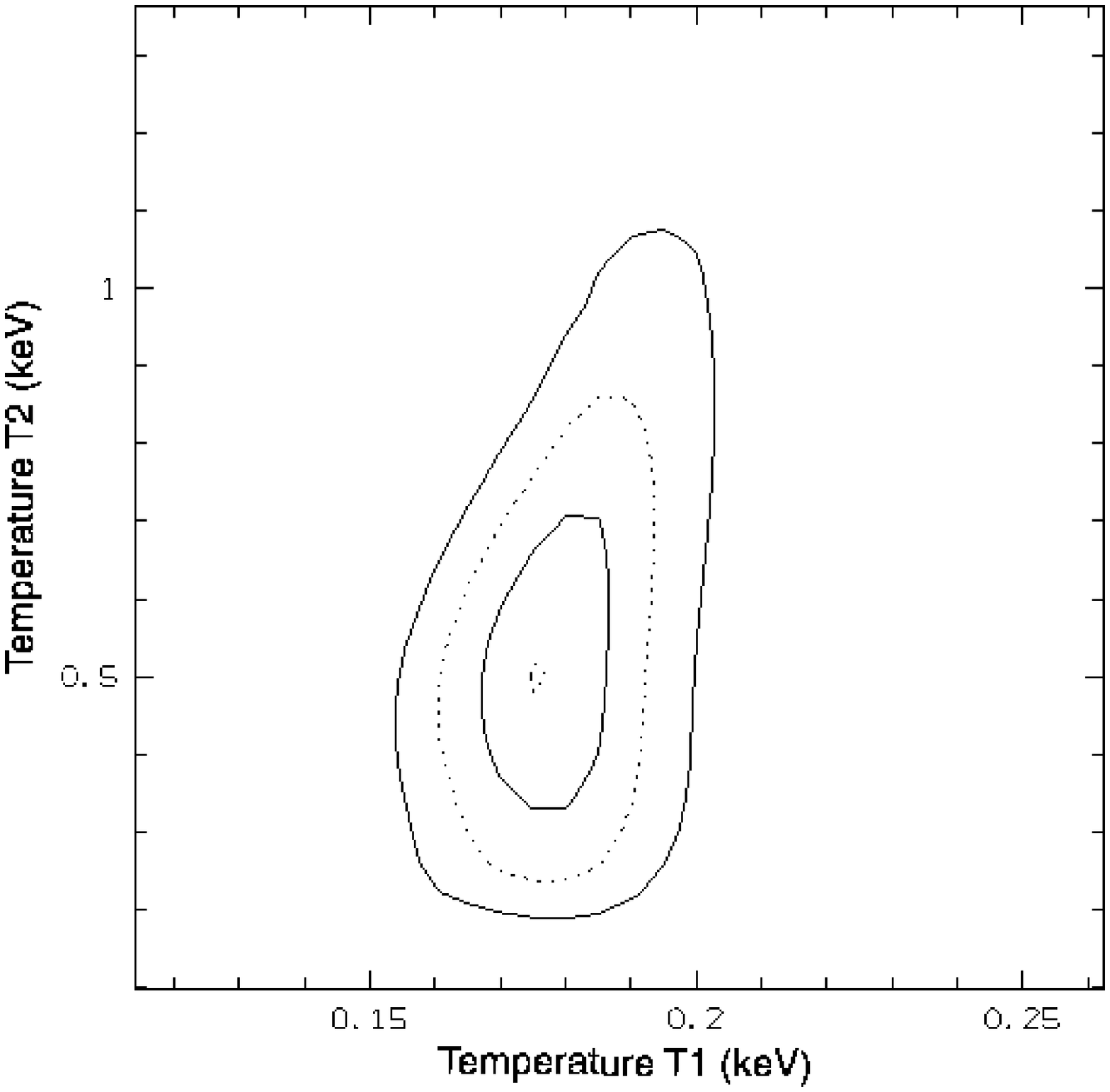,width=85mm,angle=-90}}
{\centering\psfig{figure=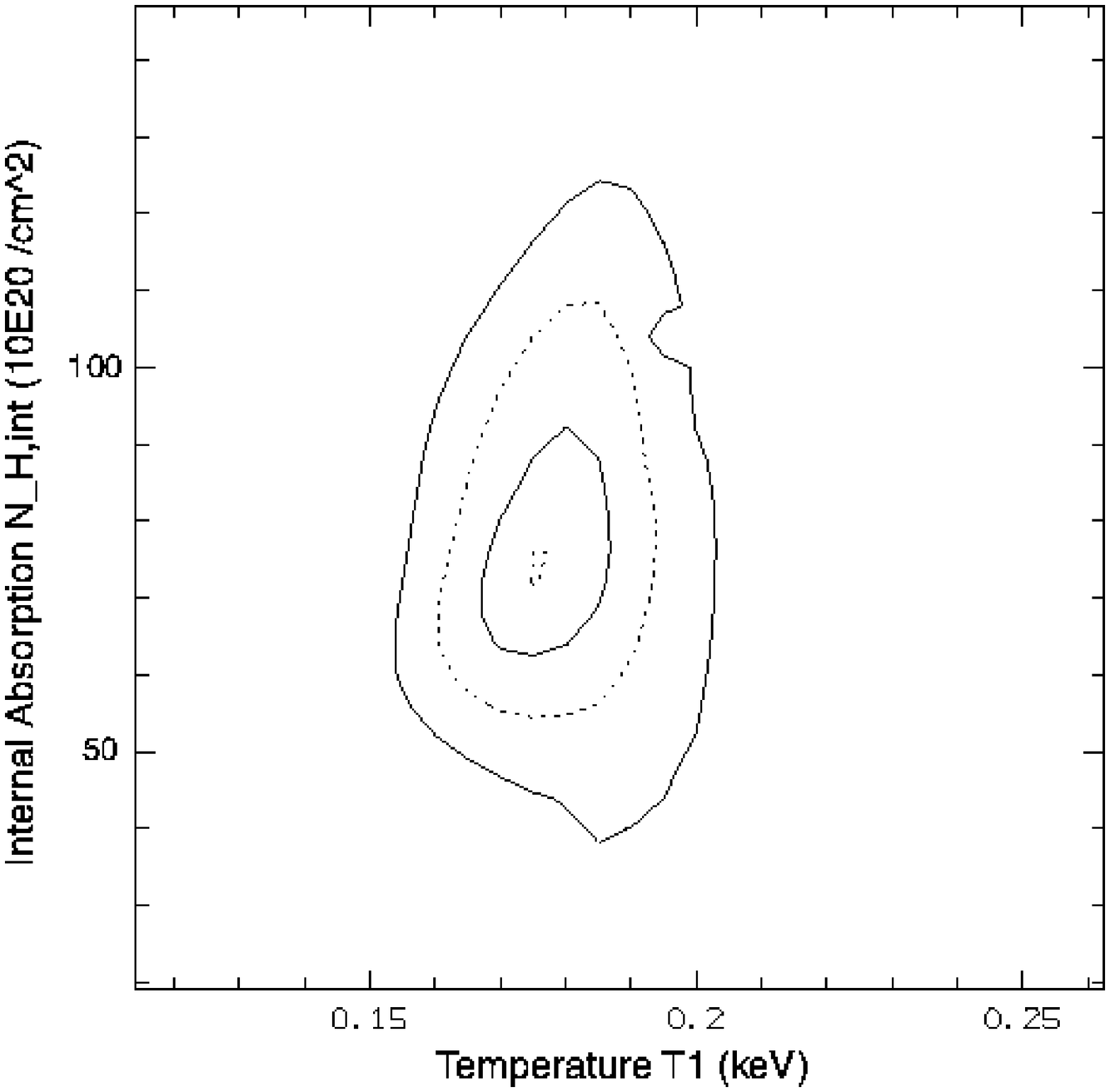,width=85mm,angle=-90}}
\caption[]{$\chi^2$ contours for the model fit (a combination of a thermal
Raymond-Smith plasma of temperature $T_1$ with an internally absorbed
thermal plasma of temperature $T_2$, both affected by Galactic foreground
absorption) to the spectrum of the extended emission of M83. The region
of the $\chi^2$ minimum is indicated by the inner dashed contour followed
by contours corresponding to 68.3\%, 96.5\% and 99.7\% confidence levels.
The number of ``interesting parameters" (cf. Avni 1976) is three. 
The {\it{left}} panel shows the plot of the parameter set $T_1$ and $T_2$, 
the {\it{right}} that of $T_1$ and $N_{\rm H}^{\rm int}$}
\label{restconf}
\end{figure}
 
Again the best-fit is obtained by combining a soft X-ray emitting warm
($0.18\pm^{0.02}_{0.01}$~keV$\cor2.1\cdot10^6$~K) plasma with an 
internally absorbed hot ($0.49\pm^{0.23}_{0.16}$~keV$\cor5.7\cdot10^6$~K) 
gas (see Table~\ref{spektrum} and Fig.~\ref{specrest}). 
Figure~\ref{restconf} shows the confidence contours for this spectral fit. 
The X-ray luminosity of the M83 extended diffuse emission in the 
{\it broad} \ROSAT band then is $L_{\rm x}=3.6\cdot10^{40}$~erg s$^{-1}$ 
(75\% due to the warm and 25\% due to the hot gas component).
The X-ray luminosity in the soft energy band, computed for the
best-fit parameters, is $L_{\rm x}({\rm soft})=1.4\cdot10^{40}$~erg
s$^{-1}$ and originates almost completely from the cooler gas component,
as it is the case for the nuclear area (see above).
  
\section{Discussion}
\label{discu}

\subsection{Point-like sources}
\label{discu_point}
We detected 13 point-like sources in M83, 10 of which were previously 
unknown, down to a limiting sensitivity of $\sim 0.8\cdot10^{38}$~erg s$^{-1}$.
Four of the seven \EINSTEIN detected sources could not be found again and are
probably time-variable X-ray binaries. 
The number of background sources in the X-ray emitting area of M83
at our limiting luminosity is expected to be only 1.2, based
on the luminosity function of sources from the \ROSAT
medium-sensitivity survey (Hasinger et al. 1991). We note that this 
value represents an upper limit as the absorbing material in the disk of 
M83 further reduces the chance to detect background sources.

Eight of the detected X-ray sources in M83 are positionally correlated 
with \HII regions and/or \HI voids and could be giant bubbles of hot gas 
in star-forming regions, young supernova remnants (Marston et al. 1995),
ultraluminous X-ray binaries or a combination of all these
components (Williams \& Chu 1995). Such a coincidence of X-ray sources 
and \HI holes has been detected in the face-on galaxy M51 
(Ehle et al. 1995) and M101 (Snowden \& Pietsch 1995, Williams \& Chu 1995).
Kamphuis et al. (1991) showed for one of the holes in M101 that it is indeed
associated with an expanding \HI shell. Diffuse X-ray emission from hot
gas in \HI shells also was found within the LMC (Chu \& Mac Low 1990;
Wang \& Helfand 1991; Bomans et al. 1994). The resolved shells in the
LMC, however, only reach X-ray luminosities of about $10^{37}$~erg~s$^{-1}$,
factors of more than 10 less than the one required for the M83 point
sources. This indicates that hot gas confined in a giant bubble can not
explain the high X-ray luminosity of these sources.
A more detailed discussion of time variability, properties of
individual sources and a comparison with \HI channel maps
should make use of sensitive \ROSAT HRI observations 
of M83 (Blair, in prep.) which allow a more reliable source detection 
and separation of point sources from the surrounding diffuse emission.

Our PSPC observations of M83 allow a crude estimate of the amount of
undetected point-like sources contributing to the luminosity of the
extended emission (calculated in Sect.~\ref{res_diff}). As the number 
of detected sources in M83 is rather low, the integral luminosity 
distribution of sources in M31 (a single power-law with
slope -0.8; Trinchieri \& Fabbiano 1991) is assumed. Furthermore, assuming that
only sources above $1\cdot10^{34}$ erg s$^{-1}$ contribute significantly to
the luminosity due to point-like sources (this cutoff is based on 
observations of X-ray binaries in the Milky Way), we integrated 
the logN-logS distribution normalized to the total luminosity 
of detected point-like sources in M83.
The luminosity due to undetected plus detected point-like sources is found
to be $1.2\cdot10^{40}$ erg s$^{-1}$, a factor of $\sim2.5$
higher than what has been detected. Undetected point-like sources
might explain $\sim 20\%$ of the extended diffuse X-ray emission.
Supper et al. (1997) found a flattening of the luminosity distribution
with decreasing luminosities for the globular cluster sources in M31. If 
this is true also for point-like sources in M83, the X-ray 
luminosity due to undetected sources would be even smaller. 
This result is in good agreement with that from \GINGA observations,
namely that a thermal component is making a considerable contribution in the
X-ray emission rather than LMXBs or X-ray pulsars (Ohashi et al. 1990).

\subsection{Nuclear region}
\label{discu_nuc}
Earlier \EINSTEIN IPC observations could not be used to investigate
the spectral properties of the extended and nuclear emission of M83
because of the short exposure time ($\sim 6000$ s) and detector gain
variations (Trinchieri et al. 1985). Spectral observations of M83 with the
\GINGA satellite (Ohashi et al. 1990) also do not allow a separation of the 
nuclear emission from
the remaining diffuse emission because of their low angular resolution.
Spectral modelling based on \ROSAT observations, however, tells us 
that the nuclear X-ray emission is complex.
The nucleus itself is detected as a bright
point-like source by our source detection algorithm and shows a
relatively hard spectrum. The \ROSAT observed luminosity of the 
nuclear region of M83 is 
$L_{\rm x}$(0.1-2.4 keV)$\sim1.6\cdot10^{40}$ erg s$^{-1}$,
in agreement with the \EINSTEIN observations within the range
of uncertainties ($L_{\rm x}$(0.5-3.0 keV)$\sim1.1\cdot10^{40}$ erg s$^{-1}$)
and can be compared with other known starburst
nuclei, such as NGC\,7714 ($L_{\rm x}$(\EINSTEIN)$=6\cdot10^{40}$ erg
s$^{-1}$; Weedman et al. 1981), NGC\,253 ($L_{\rm x}$(\EINSTEIN)$=
1.7\cdot10^{39}$ erg s$^{-1}$; Fabbiano \& Trinchieri 1984, scaled to 2.58~Mpc
distance), and NGC\,1808 ($L_{\rm x}$(\ROSAT)$=1.4\cdot10^{41}$ erg s$^{-1}$; 
Junkes et al. 1995).

A comparison of the observed spectra of the nuclear area and the 
disk (Figs.~\ref{specnuc} and \ref{specrest}) shows that in the nuclear 
region the emission of the hard component is stronger than that of the 
soft component, in contrast to the extended disk and halo emission of 
M83 where both spectral components contribute equally. 
The harder spectral properties of the emission from the nuclear area are 
also visible in the different hardness ratios (see Table~\ref{hr}).

The soft X-ray emission of the nucleus of M83 (Fig.~\ref{contours}a) 
is clearly extended and shows a ridge-like structure pointing 
approximately in the direction of the optically visible bar. 
Neininger et al. (1991) showed that the magnetic field also follows 
closely that bar structure, favouring the idea of a link between the hot gas
and magnetic fields. 

However, the reason for the strong asymmetry of the hard nuclear X-ray 
radiation (Fig.~\ref{contours}b and c) is still unclear: No such
asymmetrical emission features are detected at any other wavelengths.
We do not expect that the nuclear hard X-ray protrusion results from very hot
gas emerging from the nucleus and travelling through the relatively
dense galactic disk. Instead we prefer the idea that the asymmetry is
due to point-like sources distributed asymmetrically 
around the galactic nucleus. These sources are not resolved by the spatial
resolution of the PSPC-detector and might be connected to strong
star-forming regions. 
We will have to wait for sensitive M83 \ROSAT HRI
observations (Blair, in prep.) to study the structure and origin of the
extended nuclear X-ray emission in more detail. Part of the extended nuclear
emission of M83 might well be resolved into point-like bulge sources,
as it is the case for the nucleus of M51 (Immler 1996).

\subsection{Extended emission}
\label{discu_extended}
If we account for diffusion of cosmic rays away
from their sites of origin, the synchrotron scale length should be larger
than that of the extended X-ray emission which is in agreement with our
observations. The same result was found for the galaxy M51 (Ehle et al.
1995). The agreement of the scale length of the hard-band diffuse X-ray-
and of the thermal radio emission favours the
idea that the hard diffuse emission is due to hot gas produced by
star-forming regions in the galactic disk. This hypothesis is further
supported by the fact that the `hump' at $\sim 5$~kpc is also visible
in the optical light where it indicates the presence of active and
recent star formation (Talbot et al. 1979).

The smooth profile of the soft, diffuse X-ray emission together with 
its shallow radial descent argue for the idea that this X-ray 
emission component is related to a different distribution of emitting
sources. From observations of the \HI velocity field (Tilanus \& Allen 1993) 
and under the assumption of trailing spiral arms, it follows that the 
north-western part of the galactic disk is nearer to the observer.
Hence, the asymmetric distribution of the soft X-rays supports
the existence of a spherical gas halo: the galactic
disk is opaque to soft X-rays so that we only see the emission from the
front hemisphere. The asymmetric distribution
of surface brightness (Sect.~\ref{distribution}) 
is explainable as a projection effect caused by the
galactic inclination and absorption (cp. Pietsch et al. 1994).

The idea that most of the soft X-ray emission is due to hot gas located
in a huge halo above the disk of M83 is further supported by
the spectral characteristics of the diffuse emission:
The X-ray spectrum of the extended emission is complex and cannot
be described by simple models. This is understandable, as observing
a face-on galaxy like M83 must result in a superposition of 
different X-ray emission components from the disk and halo that are 
affected by internal absorption from the \HI gas distribution in the 
galactic disk. This statement is supported by X-ray observations of
edge-on galaxies where these components can be viewed
separately in and above the disk. We note, however, that only
observations of face-on galaxies allow the study of the distribution of X-ray
emission with respect to the (underlying) disk.
Our best fit (with a resonably small amount of free
parameters) is a two-temperature Raymond-Smith plasma with both internal
and external absorption, the latter fixed to the \HI column density
of the Milky Way in the direction to M83. A similar two-temperature
model was found to give a good description of the diffuse X-ray emission
from NGC~4631 (Wang et al. 1995). As these authors, we expect
not only two discrete temperatures but probably a continuous
temperature distribution in the real disk and halo gas. In future, more
sophisticated spectral models in combination with X-ray telescopes,
offering higher energy resolution, might allow a better description.

If the diffuse X-ray emission in the soft energy band is assumed to be
due to hot halo gas, it is possible to calculate the gas density 
$n_{\rm e}$, mass $m_{\rm
gas}$ and cooling time $\tau$ of that plasma. To this end we use the model
of thermal cooling and ionization equilibrium of Nulsen et al. (1984)
where $L_{\rm x}({\rm soft})=1.11\cdot\Lambda(T) n_{\rm
e}^2 V \eta$. The unknown filling factor $\eta$ allows for some 
clumpiness of the gas. 
For the gas temperatures of the nuclear region and
of the extended halo emission ($\sim2.1\cdot10^6$~K) Raymond et 
al. (1976) give a cooling coefficient $\Lambda(T)$ of  
$\sim10\cdot10^{-23}$ erg cm$^3$ s$^{-1}$.
To calculate the physical parameters of the hot gas component one has to make
assumptions about the emitting volume $V$: here we take into account that
the soft diffuse X-ray emission from regions behind the disk of M83 is
absorbed by the neutral hydrogen in the galactic disk. Column densities of
$\sim0.5\cdot10^{21}$~cm$^{-2}$ already lead to a nearly complete
absorption of the soft-band emission. Such and even higher $N_{\rm H}$ 
values are observed in the disk of M83 (Tilanus \& Allen 1993).
We assume that the nuclear soft X-ray radiation is emitted from a 
hemisphere with radius 1\arcmin~ (see above) and that the extended soft
emission from the halo of M83 
fills a hemisphere with radius $5\arcmin\cor13$~kpc (see above), minus a 
cylinder of radius 1\arcmin~ around the center where the emission 
clearly is dominated by the gas of the nuclear area. The calculated
gas parameters are presented in Table~\ref{gas}.
\begin{table}
\caption[]{Halo Gas Parameters for M83}
\label{gas}
\begin{flushleft}
\begin{tabular}{lccc}
\hline
\noalign{\smallskip}
&$n_{\rm e}$&$m_{\rm gas}$&$\tau$\\
Region&(cm$^{-3}$)&$(M_{\sun})$&(yr)\\
\noalign{\smallskip}
\hline
\noalign{\smallskip}
Nuclear area& $7.0\cdot10^{-3}/\sqrt{\eta}$ & $0.1\cdot10^8\cdot\sqrt{\eta}$ & $0.5\cdot10^8\cdot\sqrt{\eta}$\\
Rest of M83   & $1.2\cdot10^{-3}/\sqrt{\eta}$ & $2.1\cdot10^8\cdot\sqrt{\eta}$ & $2.8\cdot10^8\cdot\sqrt{\eta}$\\
\noalign{\smallskip}
\hline
\end{tabular}
\end{flushleft}
\end{table}

The total mass of the detected hot gas (nuclear area plus halo) is
$2.2\cdot10^8\sqrt{\eta}~M_{\sun}$ which is only $\la0.2\%$ of the total gas
mass (cp. Table~\ref{parameters}). Similarly low mass contributions are 
found for the galaxies M51 (Ehle et al. 1995) and NGC~1566 (Ehle et al.
1996). The cooling times are longer than the time scale for starburst
activity which is estimated to be around $\sim10^7$ yr (Rieke et al. 1988,
Trinchieri et al. 1985).

\subsection{Hot gas, galactic winds and magnetic fields}
From X-ray observations of the hot gas it is possible to calculate the
thermal energy density of this component of the interstellar medium:
$U_{\rm therm}=n_{\rm e}k_BT$. 
(It is, however, still not clear whether the ions are 
in thermal equilibrium with the plasma electrons (see Lesch 1990) and 
how strongly they contribute to $U_{\rm therm}$). 
A comparison of the nuclear area and
the halo region of M83 (Table~\ref{gas}) shows that $U_{\rm therm}$ is 
a factor of 6 higher in the nuclear region. Here the gas can escape much
more easily from the galactic disk and rise up into the halo, e.g. in
the form of a galactic wind channeled by vertical magnetic fields
(Breitschwerdt \& Schmutzler 1994). High Faraday rotation measures and
strong depolarization of polarized radio emission in the central part of
M83 support this concept (Neininger et al. 1991, Ehle 1995).
Further evidence of nuclear gas outflow comes from the detection of
absorption lines blueshifted by 1000 km s$^{-1}$ relatively to the system's
velocity (Bohlin et al. 1983).

\begin{figure}
{\centering\psfig{figure=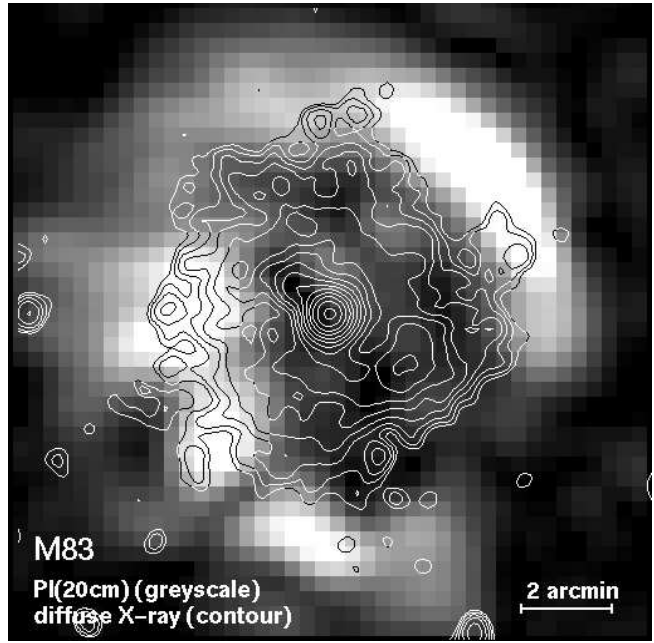,bbllx=70pt,bblly=300pt,bburx=520pt,bbury=754pt,width=85mm}}
\caption[]{Broad-band \ROSAT PSPC contours of the diffuse X-ray emission
of M83 (detected point sources are subtracted), overlaid onto a greyscale
map of the distribution of polarized intensity at $\lambda20$~cm (cp.
Sukumar \& Allen 1989). Contours are $(1,1.4,2,2.8,4,...)\times1\cdot10^{-3}$
cts s$^{-1}$ arcmin$^{-2}$ above the background}
\label{depol}
\end{figure}

Whereas the electron densities of the hot gas (as derived from the
presented X-ray data) are in general
too low to explain the observed depolarization (see below), the density
in the nuclear area is relatively high and may contribute to the Faraday
effects. A comparison of the polarized radio emission at $\lambda20$~cm
and the broad-band diffuse X-ray emission (Fig.~\ref{depol}) shows indeed a
correlation of strong X-ray emission with low polarized intensities in the
central area of M83.
This may indicate that the diffuse X-ray emitting
hot gas in the nuclear area depolarizes by the effect of
internal Faraday dispersion along the line of sight, produced by
turbulent magnetic fields and thermal gas inside the emitting source (Burn
1966, Ehle \& Beck 1993). Adopting a strength of
the turbulent magnetic field of $25~\mu$G (Neininger et al. 1993), a
pathlength of $1\arcmin\cor2.6$~kpc (see above) and a typical turbulence
scale length of 50~pc requires electron densities of
$\sim2\cdot10^{-2}/\sqrt{\eta}$~cm$^{-3}$
to explain the depolarization of $DP\sim0.02$ at $\lambda20$~cm
in the central region of M83. This gas density is comparable, within the
range of uncertainties, to that calculated from our X-ray observations
(see Table~\ref{gas}).
 
The electron densities of thermal gas in the galactic halo required to
explain the strong depolarization outside the nuclear area are in
the range of $n_e\ga0.03$~cm$^{-3}$ (e.g. Ehle 1995).
These values are {\it{inconsistent}} with the halo densities as derived from
our \ROSAT X-ray observations. Either Faraday depolarization occurs in
cooler ionized gas not radiating in X-rays (warm absorber), or the
turbulence scale length in the halo is much larger than in the disk, or the
assumption of ionization equilibrium (Raymond-Smith plasma with one
specific temperature) is not valid. The latter possibility is supported
by the model of galactic winds and delayed recombination
(Breitschwerdt \& Schmutzler 1994). In their simulation the soft X-ray
emission can be explained by a cooler (typical $10^4$~K) gas which might
have electron densities that are larger by a factor 3-5
(Schmutzler, priv. comm.) than the values given in Table~\ref{gas}.
 
From radio polarization observations it is possible to compare $U_{\rm
therm}$ with the energy density of the magnetic field:
$U_{\rm magn}=B^2/8\pi$. Assuming equipartition between cosmic rays and
magnetic fields, Neininger et al. (1993) calculated, for a synchrotron
emitting disk of $1.0\pm0.5$~kpc thickness, a total magnetic field
strength of $B=11\pm2~\mu$G for M83. 
To scale the disk field strength to that of the
halo we use observations of the edge-on galaxy NGC~253 (Beck et al. 1994)
where the field strength at a height of $z\sim2.5$~kpc above the galactic
disk is found to be about 80\% of the disk value.
 
Allowing the filling factor to vary between 0.1 and 0.8 and including a
0.1 keV uncertainty for the gas temperature, we obtain a rough estimate
of the plasma parameters $\beta=U_{\rm therm}/U_{\rm magn}$ of
$0.2\pm0.1$ both for the nuclear area and for the halo. The galactic
wind model (see above) leads to an even lower $\beta$.
  
Although there are other contributions to the total
energy density of the hot interstellar medium (e.g. turbulent
motions, cosmic rays) it is obvious that the magnetic field plays an
important role for galactic haloes. How it might channel galactic
winds, support or hinder the escape of hot gas from the galactic disk,
or heat the gas via MHD waves or magnetic reconnection is still under
discussion.
   
\subsection{Comparison with other galaxies}
 
Assuming that most of the mechanical energy supplied to the
interstellar gas is from supernovae, one might ask whether
this energy input is high enough to explain the thermal energy
of the hot X-ray emitting gas (cf. Fabbiano et al. 1997). Based on an
idealized model for the X-ray emission (a single spherical expanding
bubble, cf. Heckman et al. 1996), the supernova rate of M83
(0.29 - 0.49~yr$^{-1}$; see below) might produce a
thermal energy of $(1.3 - 2.2)\cdot 10^{56}$~erg integrated over a
typical starburst period of $10^7$~yr. This value is
indeed above the thermal energy of the hot gas in the halo region of
$\sim0.4\cdot 10^{56}\sqrt{\eta}$~erg, using the same assumptions for the
volume of the emitting halo and the soft-band X-ray luminosity as in
Sect.~\ref{discu_extended}. This agreement between
model and observations supports the hypothesis that the diffuse X-ray
emission arises from hot gas originally heated by supernovae.
 
The luminosity of the diffuse X-ray emission in M83 of
$3.6\cdot10^{40}$~erg s$^{-1}$ in the broad \ROSAT energy band is very
high. Even if one assumes that only the soft X-ray emission of
$1.8\cdot10^{40}$~erg s$^{-1}$ of the extended and nuclear component
originates from a halo above the galactic disk (Sect.~\ref{discu_extended})
this component is still much brighter than the halo emission of the
edge-on galaxies NGC~4631 ($0.1-0.2\cdot10^{40}$~erg s$^{-1}$;
Vogler \& Pietsch 1996) or NGC~253, a galaxy well known for its
huge X-ray halo ($\sim 0.17\cdot10^{40}$~erg s$^{-1}$; Pietsch et al. 1997).
 
Not only is the absolute value of the X-ray luminosity of the diffuse
component in M83 high but it also strongly dominates the total emission.
Even if we take into account that 20\% of the so-called diffuse emission
from the disk area could in fact be point-like sources below the
detection-limit of $0.8\cdot10^{40}$ erg s$^{-1}$ (Sect.~\ref{discu_point}),
the fraction of diffuse emission in M83 is 78\% of the total
luminosity, comparable to the corresponding value of 70\% for NGC~253 (see
Table~\ref{vergleich}). Values for NGC~253 are based on results of the
data analysis by Pietsch et al. (1997) and Vogler et al. (1997) for an
assumed distance of 2.58~Mpc.
 
\begin{table}
\caption{Comparison of X-ray luminosities of emission components in M83 and NGC~253}
\label{vergleich}
\begin{flushleft}
\begin{tabular}{ccccc}
\hline
\noalign{\smallskip}
luminosity in $10^{40}$ erg s$^{-1}$&\multicolumn{2}{c}{M83}&\multicolumn{2}{c}{
NGC~253}\\
\noalign{\smallskip}
\hline
\noalign{\smallskip}
total                            & 5.7 & 100\% & 0.46$^{\rm a}$ & 100\% \\
\noalign{\smallskip}
\hline
\noalign{\smallskip}
halo $^{\rm b}$                  & 1.8 &  32\% & 0.17 &  37\% \\
disk + bulge, diffuse $^{\rm c}$ & 2.7 &  47\% & 0.15 &  33\% \\
disk + bulge, point-like sources & 1.2$^{\rm d}$ &  21\% & 0.14 &  30\% \\
\noalign{\smallskip}
\hline
\end{tabular}
\end{flushleft}
$^{\rm a}$ - only half of the total halo emission is included
for comparison with the face-on galaxy M83\\
$^{\rm b}$ - assuming that the soft diffuse emission originates from the halo\\
$^{\rm c}$ - the remaining part of the diffuse emission is assumed to be
present in the disk and bulge. For M83 it is reduced by 20\% to account for
undetected point-like sources\\
$^{\rm d}$ - for M83 we give the sum of detected plus estimated undetected 
point-like sources \\
\end{table}

Although the luminosities of the emission components in M83 are much higher
than that of NGC~253 - a fact partly explainable by the still uncertain
extragalactic distances - there is an agreement of the relative
contributions to the total luminosity. The (distance-independent) mean 
surface brightness $I_x$ of the soft diffuse emission of M83 is about 
twice that of NGC~253 (Table~\ref{input}). In edge-on view the X-ray halo 
of M83 would probably look even more spectacular than that of NGC~253.
 
To search for a common reason for this difference between M83 and NGC~253
we compare these two galaxies with other galaxies where
diffuse X-ray emission is detected. We do this with respect to their
supernova rates and discuss also distance-independent parameters like
the X-ray surface brightnesses $I_{\rm x}$ and the energy input per
surface area, $\dot{E}^{\rm tot}_{\rm A}$, a quantity
that measures the activity in the galactic disk and seems to be important
for the evolution of galactic radio haloes (Dahlem et al. 1995,
Dumke et al. 1995, Golla 1997).
Results for four galaxies are presented in Table~\ref{input}.
\begin{table}
\caption{Comparison of M83 and other galaxies with diffuse X-ray emission}
\label{input}
\begin{flushleft}
\begin{tabular}{ccccccc}
\hline
\noalign{\smallskip}
Galaxy & $D$ & $r_{\rm SF}$ & $\nu_{\rm SN}^{\rm nt}$&$\nu_{\rm SN}^{\rm FIR}$
       & $\dot{E}^{\rm tot}_{\rm A}$ & $I_{\rm x}$\\
  (1)  & (2) &     (3)      &           (4)          &          (5)
       &           (6)               &  (7)       \\
\noalign{\smallskip}
\hline
\noalign{\smallskip}
M83     & 8.9  & 9.9  & 0.29 & 0.49 & 3.2  & 2.9 \\
NGC~253 & 2.58 & 7.7  & 0.11 & 0.15 & 1.9  & 1.5 \\
	&      &2.3$^{\dagger}$&&   & 22.7 &     \\
NGC~3628& 6.7  & 7.0  & 0.04 & 0.06 & 0.9  & 1.0 \\
NGC~4565& 9.7  & 16.9 & 0.02 & 0.04 & 0.1  & 0.5 \\
\noalign{\smallskip}
\hline
\end{tabular}
\end{flushleft}
$\dagger$ - in NGC~253 most of the star formation is limited to the
inner disk (Scoville et al. 1985: extended inner stellar disk)\\
(2) - assumed distance in Mpc\\
(3) - radial extent of star-forming disk in kpc, estimated from IRAS CPC
observations (van Driel et al. 1993)\\
(4) - supernova rate in yr$^{-1}$, calculated from the power of nonthermal
radio emission using formulae given by Dahlem et al. (1995)\\
(5) - supernova rate in yr$^{-1}$, calculated from FIR luminosities given
by Young et al. (1989)\\
(6) - mean energy input rate per unit surface area (as defined in Dahlem
et al. 1995) in $10^{-3}$~erg~s$^{-1}$~cm$^{-2}$ \\
(7) - X-ray surface brightness in $10^{-6}$~erg~s$^{-1}$~cm$^{-2}$,
calculated over the area of soft halo emission in NGC~253, NGC~3628 and 
NGC~4565 and over the extent of the soft diffuse emission in M83, 
respectively.\\
\end{table}
	 
The supernova rates and consequently the energy input rates
into the interstellar medium are well correlated
with the X-ray surface brightnesses. The energy input rate
in case of M83 is very high and even higher than that for NGC~253
(if averaged over the whole area of star formation).
The energy input rate from the inner star-forming disk of
NGC~253 alone is by far the highest, a result that might explain
the extreme vertical extent of the X-ray halo emission of that galaxy.
 
While there is increasing evidence that star formation activity in the disks
of spiral galaxies is related to the existence and structure of radio haloes,
further studies are needed to test the correlation between the energy
input by supernovae in the disk and the surface brightness and
evolution of X-ray haloes.
Future X-ray satellites, with their higher spatial resolution and
sensitivity especially for the soft X-ray component, will help to
disentangle the emission of hot gas in haloes and disks.
 
An important finding of our study is that
outflow models for the transport of hot gas up into the halo should not
be restricted to nuclear starbursts and superwinds (Suchkov et al. 1994),
but should include the whole star-forming galactic disk.
M83 is a good example of a X-ray halo produced by an active star-forming
disk.
\begin{acknowledgements}We would like to thank David Malin, AAO, for
providing the optical image on M83 and Stuart Ryder, Uni. of NSW, for
the H$\alpha$ map. J\"urgen Kerp, RAIUB, is
acknowledged for helpful comments on the manuscript as well as Andreas Vogler,
MPE, for his help during the data reduction. 
Our anonymous referee is acknowledged for numerous helpful suggestions.
The work of ME was supported by the {\it Deutsche Agentur f\"ur 
Raumfahrtangelegenheiten (DARA)} project number 50 OR 9206 and 50 OR 9405
and the {\it Deutsche Forschungsgemeinschaft (DFG)} grant number Eh 154/1-1.
The \ROSAT project is supported by the German {\it Bundesministerium f\"ur 
Bildung, Wissenschaft, Forschung und Technologie} and the 
{\it Max-Planck-Gesellschaft}. 
\end{acknowledgements}

%
%

%
\end{document}